\documentclass[12pt]{article}
\usepackage[mathscr]{eucal}
\usepackage{amsmath,amsfonts,amssymb,amsthm}
\usepackage[hypertex]{hyperref}
\newcommand{\comm}[1]{}

\advance\textheight\topskip
\numberwithin{equation}{section} \makeatletter
\@addtoreset{equation}{section}

\renewcommand{\tilde}{\widetilde}
\renewcommand{\hat}{\widehat}

\newcommand{\bref}[1]{\textbf{\ref{#1}}}

\newcommand{\p}[1]{|#1|}
\newcommand{\gh}[1]{\mathrm{gh}(#1)}

\newcommand{\dd}{\partial}
\renewcommand{\d}{\partial}

\def\ba{\begin{array}}
\def\ea{\end{array}}
\def\dps{\displaystyle}

\renewcommand{\geq}{\,{\geqslant}\,}
\renewcommand{\leq}{\,{\leqslant}\,}

\newcommand{\inner}[2]{\langle #1{,}\,#2\rangle}
\newcommand{\binner}[2]{%
  {\langle}\kern-4.15pt{\langle}#1{,}\,#2{\rangle}\kern-4.15pt{\rangle}}
\newcommand{\commut}[2]{[#1{,}\,#2]}

\newcommand{\pb}[2]{\left\{{}#1{},{}#2{}\right\}}

\newcommand{\half}{\mathchoice{%
    \ffrac{1}{2}}{\frac{1}{2}}{\frac{1}{2}}{\frac{1}{2}}}

\newcommand{\ffrac}[2]{\raisebox{.5pt}%
  {\footnotesize$\displaystyle\frac{#1}{#2}$}\kern1pt}

\newcommand{\derham}{\boldsymbol{d}}

\newcommand{\dl}[1]{\mathchoice{\ffrac{\dd}{\dd #1}}{\frac{\dd}{\dd
      #1}}{\ffrac{\dd}{\dd #1}}{\ffrac{\dd}{\dd #1}}}

\newcommand{\st}[2]{{\overset{#1}{#2}}}

\newcommand{\ddl}[2]{\ffrac{\dd #1}{\dd #2}}

\def\const{\mathop\mathrm{const}\nolimits}

\newcommand{\manifold}[1]{\mathscr{#1}}
\newcommand{\manX}{\manifold{X}}

\newcommand{\manM}{\manifold{M}}

\newcommand{\Liealg}{\mathfrak}
\newcommand{\algg}{\Liealg{g}}

\def\cC{\mathcal{C}}

\def\cE{\mathcal{E}}
\def\cF{\mathcal{F}}

\def\cK{\mathcal{K}}
\def\cL{\mathcal{L}}

\def\cN{\mathcal{N}}

\def\cP{\mathcal{P}}

\def\cT{\mathcal{T}}

\newcommand{\ee}{{e}}
\newcommand{\psymp}{{\chi}}
\newcommand{\dm}{{d}}
\newcommand{\Vol}{\mathcal{V}}
\newcommand{\SM}{L}
\newcommand{\HM}{H_{\manM}}
\newcommand{\Lie}{\cL}
\newcommand{\eom}{\cE}

\def\BGnlp{Barnich:2010sw}
\def\BG-Poincare{Barnich:2009jy}
\def\Fedosov-book{Fedosov:1996fu}

\begin{document}

\begin{flushright}
FIAN-TD-2013-21 \\
\end{flushright}

\begin{center}

{\Large\textbf{
Frame-like Lagrangians   and   
presymplectic \\ \vspace{3mm}
AKSZ-type sigma models}}

\vspace{.9cm}

{\large Konstantin Alkalaev$^a$ and  Maxim Grigoriev$^{a,b}$}

\vspace{0.5cm}

\textit{$^{a}$I.E. Tamm Department of Theoretical Physics, \\P.N. Lebedev Physical
Institute,\\ Leninsky ave. 53, 119991 Moscow, Russia}

\vspace{0.5cm}

\textit{$^{b}$Moscow Institute of Physics and Technology, Dolgoprudnyi,\\
141700 Moscow region, Russia}

\vspace{0.5cm}

\thispagestyle{empty}


\end{center}

\begin{abstract}

We study supergeometric structures underlying frame-like
Lagrangians. We show that for the theory in  $n$ spacetime
dimensions both the frame-like Lagrangian and its gauge symmetries
are encoded in the target supermanifold equipped with the odd vector
field, the closed $2$-form of ghost degree $n-1$, and the scalar potential
of ghost degree $n$. These structures satisfy
a set of compatibility conditions ensuring the gauge invariance of
the theory. The Lagrangian and the gauge symmetries have the same
structures as those of AKSZ sigma model so that frame-like
formulation can be seen as its presymplectic generalization.
In contrast to the conventional AKSZ model the generalization allows to describe systems with
local degrees of freedom in terms of finite-dimensional target space. We argue that
the proposed frame-like approach is directly related de Donder--Weyl polymomentum Hamiltonian formalism.
Along with the standard field-theoretical examples like Einstein--Yang--Mills theory we consider free higher spin fields, multi-frame gravity, and parameterized systems. In particular, we propose the frame-like action for free totally
symmetric massless fields that involves all higher spin connections on an equal footing.

%

\end{abstract}

\vspace{1cm}

\newpage

%

\section{Introduction}


Manifestly diffeomorphism-invariant formulations in terms of
$p$-forms are extremely useful in the context of
supergravity~\cite{D'Auria:1982pm,DAuria:1982nx} and higher spin
gauge
theories~\cite{Vasiliev:1980as,Vasiliev:1988xc,Lopatin:1988hz,Zinoviev:2003ix,Alkalaev:2003qv,
Skvortsov:2008sh, Boulanger:2012bj,Vasilev:2011xf} (for a review
see~\cite{Bekaert:2005vh}). One can usually assume the Lagrangian
to be first order and to involve only the wedge product and de
Rham differential. The structure of the respective action is
remarkably similar to the standard extended Hamiltonian action
while its gauge symmetries can be expressed through an associated
BRST-type differential.

In this work we study the supergeometric structures underlying
first order frame-like Lagrangians and propose a set of
compatibility conditions for them which ensure the gauge
invariance. More precisely, the basic object is the graded
supermanifold, the target space $\manM$, where the $p$-form fields
take values so that the degree of a coordinate is a form degree of the
respective field. The target space comes equipped with a
presymplectic potential $1$-form $\chi$, an odd vector field $Q$
of degree one, and a function $\HM$ which is an analog of the
usual Hamiltonian. This data is enough to formulate a first order
Lagrangian which is by construction invariant under the
diffeomorphisms and the extra gauge transformations generated by
the odd vector field provided some natural compatibility
conditions are satisfied.

In the special  case where the presymplectic form is invertible the equations of motion take the form
of a free differential algebra and the compatibility conditions require the odd vector field to be nilpotent.
In this case the action can be reformulated as the so-called AKSZ sigma model~\cite{Alexandrov:1995kv}
(for further developments and applications see, \textit{e.g.},~\cite{Cattaneo:1999fm, Grigoriev:1999qz,Batalin:2001fc, Park:2000au,Roytenberg:2002nu,Kazinski:2005eb,Barnich:2009jy}) whose characteristic feature is that its Batalin-Vilkovisky
(BV) structure is manifest already at the level of the classical action.

Although it is well-known that if the number of fields is finite
and $n>1$ the AKSZ sigma model is necessarily topological, the
first order frame-like Lagrangians studied in this paper are
deeply related to AKSZ sigma models and the Batalin-Vilkovisky
approach. Namely, at the level of equations of motion any gauge
theory can be represented as an AKSZ sigma model with infinite
number of fields (the so-called parent formulation) by adding
generalized auxiliary fields and (if necessary)
parametrization~\cite{\BGnlp}.

The Lagrangian counterpart of the parent formulation is also
known~\cite{Grigoriev:2010ic}. In this case however the interpretation
of the resulting theory is more subtle. One either needs
to truncate the resulting formulation or to carefully define
the space of allowed field configurations~\cite{Grigoriev:2012xg}. It was shown that
various frame-like actions can be systematically obtained by equivalent reduction of the parent formulation \cite{Grigoriev:2010ic,Grigoriev:2012xg}. Moreover, from this perspective all the structures
of the reduced theory originate from the BV structure of the parent formulation.
In particular, the canonical BV odd symplectic potential gives rise to the presymplectic
form $\chi$, while the BRST differential determines the $Q$ odd vector field upon reduction.

First order frame-like action can be considered as a multidimensional generalization
of the conventional extended Hamiltonian action. Moreover, in many cases the above supermanifold, presymplectic potential
and the generalized Hamiltonian directly give rise to the polymomentum phase space, the canonical $n$-form, and
the Hamiltonian of the de Donder--Weyl polymomentum formulation of the theory~(see, \textit{e.g.},~\cite{Gotay:1997eg,Kanatchikov:1997wp,Kanatchikov:2000jz}).  Although straightforward for
theories without gauge-invariance or relatively simple models like Yang-Mills theory the identification
is not so natural as far as genuine diffeomorphism invariance like that of Einstein gravity or Chern-Simons model
comes into the game. In this case it seems that the target space supermanifold  itself plays a more fundamental role
and is to be interpreted as a proper version of the polymomentum phase space.

The paper is organized as follows. In Section \bref{sec:framelike} we discuss general properties of frame-like
Lagrangians, starting form the usual AKSZ construction in Section \bref{sec:ACSZ}.
In particular, we introduce basic structures of the  AKSZ sigma model and
discuss gauge symmetries of its Lagrangian. In Section \bref{sec:relax} we
consider possible generalizations of AKSZ sigma models. To this end, we introduce
basic geometric structures on the target manifold and propose a set of their
compatibility conditions  that allow one to build a gauge invariant frame-like Lagrangian.
In Section \bref{sec:examples} the approach is illustrated by a number
of field-theoretical examples. In particular, in Sections from \bref{sec:gravity} to \bref{sec:linearized}
we reformulate various gravity models, including their multi-frame generalizations and higher spin frame-like Lagrangians.
In Section \bref{sec:poly} we extend our discussion to generic one-dimensional constrained Hamiltonian systems, and the so-called parameterized systems. A relation of the presymplectic AKSZ-type formulation to the
de Donder--Weyl polymomentum formulation is established in Section \bref{sec:polurel}.
Our notation and conventions, along with some basic facts in supergeometry  are collected in Appendix \bref{sec:appendix}.

\section{First order frame-like Lagrangians}
\label{sec:framelike}

\subsection{AKSZ sigma models}
\label{sec:ACSZ}

The AKSZ sigma model
is defined in terms of two supermanifolds: the target space
$\manM$ and the source $\manX$. Target $\manM$ is equipped with odd
nilpotent vector field $Q$ and the ghost degree $\gh{\cdot}$ such that
$\gh{Q}=1$. If coordinates on $\manM$ are $\Psi^A$, $A = 1,..., \dim \manM $,
then $Q = Q^{A}(\Psi)\,\d_A$, where $\d_A = \dps\dl{\Psi^A}$ (see Appendix \bref{sec:appendix}).

Source $\manX$ is equipped with odd nilpotent vector
field $\derham$ and a ghost degree also denoted by $\gh$,
$\gh{\derham}=1$. Usually, $\manX$ is taken to be odd tangent
bundle $T[1]X$ over a manifold $\manX$, \textit{i.e.} with inverse parity of the fibers.
If $x^\mu$ $\mu=1,\ldots, n=\dim X$  are coordinates on $X$
and $\theta^\mu$ with $\gh{\theta^\mu}=1$ on the fibers, then
$\dps\derham=\theta^\mu \dl{x^\mu}$ is the de Rham differential.

This data is enough to define AKSZ model at the level of equations
of motion. The fields are ghost degree zero maps from $\manX$ to
$\manM$. To each target space coordinate $\Psi^A$
of ghost degree $p$, $p\geq 0$ one associates a $p$-form field
$\dps \Psi^A(x,\theta)=\frac{1}{p!}\,\Psi^A_{\mu_1 \dots \mu_p}(x)\,
\theta^{\mu_1}\ldots \theta^{\mu_p}$, where $\gh{\Psi^A_{\mu_1 \dots \mu_p}(x)} = 0$, while coordinates with $p<0$
do not have associated fields. The equations of motion are
\begin{equation}
\label{eom-aksz}
\derham \Psi^A+Q^A(\Psi)=0\quad  \gh{\Psi^A}\geq 0\,,
\qquad\;\;
Q^A(\Psi)=0 \quad \gh{\Psi^A}<0\;.
\end{equation}
Note that in the expressions of $Q^A(\Psi)$
fields $\Psi^A(x,\theta)$ associated with negative
ghost degree coordinates should be put to zero. To avoid confusions one would
better use different conventions for fields and target
coordinates. However, in what follows we almost always assume that
negative degree coordinates are absent. In this case, the
second equation in~\eqref{eom-aksz} is missing and the equations
of motion take the form of free differential algebra~\cite{Sullivan:1977fk,DAuria:1982nx,D'Auria:1982pm}, also known as the unfolded formulation~\cite{Vasiliev:1988sa,Vasiliev:2005zu}.

The nontrivial property of AKSZ model is that its BV-BRST
description is encoded in the same geometric structures.
Fields of nonzero ghost degree can be
introduced  by simply taking a generic map from $\manX$ to
$\manM$. In terms of components, one considers
$\dps\tilde\Psi^A(x, \theta)=\sum^n_{l=1}
\frac{1}{l!}\,\st{l}{\Psi}{}^{A}_{\mu_1 \dots \mu_l}(x)\,
\theta^{\mu_1}\ldots \theta^{\mu_l}$, where
$\gh{\st{l}{\Psi}{}^{A}_{\mu_1 \dots \mu_l}}=\gh{\Psi^A}-l$, so
that at $l=\gh{\Psi^A}$ one finds ghost degree zero  fields  interpreted
as fields of the original AKSZ  model. Furthermore, the BRST differential on the
BRST extended space of fields is determined in terms of $\derham$
and $Q$. Leaving aside technical details the BRST differential is
determined by $s\tilde\Psi^A(x,\theta)=\derham
\tilde\Psi^A(x,\theta)+Q^A[\tilde\Psi(x,\theta)]$. In particular,
the equations of motion~\eqref{eom-aksz} are just
$s\tilde\Psi^A(x,\theta)=0$, where all fields of nonzero degree are
put to zero.

In the same way the gauge symmetries are encoded in
$s$ through $\delta \Psi^A=s\tilde\Psi^A$, where in the RHS one
puts to zero all the components in nonzero degree except for
degree one fields which are to be replaced by gauge parameters so
that
\begin{equation}
\label{gs-aksz}
 \delta_\lambda \Psi^A=\derham \lambda^A -\lambda^C\d_C Q^A\;,
\end{equation}
where $\lambda^A$ is a gauge parameter associated to $\Psi^A$. More precisely, if $\gh{\Psi^A}=p$ then
$\lambda^A$ is a $(p-1)$-form. Note that
coordinates with $\gh{\Psi^A}\leq 0$ do not give rise to gauge parameters.

To describe Lagrangian systems one assumes that $\manM$ is graded symplectic. More precisely,
$\manM$ is additionally equipped with a nondegenerate 2-form $\sigma_{AB}$ of ghost degree $n-1$
invariant with respect to odd vector field $Q$, \textit{i.e.},  $\Lie_Q \sigma=0$.
Suppose now that $\chi_A$ is a symplectic potential and $L$ is a Hamiltonian of $Q$, so that
$\sigma= d \chi$  and $d\SM=i_Q \sigma$, where $d$ denotes the de Rham differential on $\manM$ (see Appendix \bref{sec:appendix}). Then,
one can define the action as
\begin{equation}
\label{masteraction}
S[\Psi]=\int_{X} \big(\derham \Psi^A \chi_A (\Psi)+\SM(\Psi)\big)\,.
\end{equation}
As before, replacing ghost degree zero fields $\Psi^A(x,\theta)$ with the complete multiplet $\tilde\Psi^A(x, \theta)$ in the expression for $S[\Psi]$ results in the Batalin--Vilkovisky master action $S[\tilde \Psi]$, while the symplectic structure gives
rise to the antibracket on the space of all component fields entering $\tilde\Psi^A(x,\theta)$.

It is instructive to check gauge invariance of the above action.
To begin with we can assume that $\sigma_{AB}$ is constant by
virtue of the  Darboux theorem. In this case, the invariance of
action \eqref{masteraction} under \eqref{gs-aksz} is
straightforward. To perform this check in generic coordinates we
recall the following simple observation: suppose that under the
transformation $\delta_\lambda $ the action transforms as
$\delta_\lambda S=M^{IJ}[\Psi,\lambda]
\ddl{S}{\Psi^J}\ddl{S}{\Psi^I} $, where $M^{IJ}$ are some
functions of fields and gauge parameters (here, for simplicity, we use condensed notation). In other words, the action is invariant modulo terms of order 2 in the equations of motion.
Taking $M^{IJ}$ (graded) antisymmetric results in the  so-called trivial gauge transformations which
automatically preserve the action~(see, \textit{e.g.},~\cite{HT-book}).

Although the transformation is not a symmetry of the action, it coincides with the symmetry transformation
modulo terms vanishing on-shell. Indeed, adjusting the transformation as
\begin{equation}
\delta^\prime_\lambda \Psi^I=\delta_\lambda \Psi^I-M^{IJ} \ddl{S}{\Psi^J}\;,
\end{equation}
one finds a genuine symmetry. This simple observation is useful in the sequel
as it simplifies the description of gauge symmetries. It allows to only check the invariance modulo terms
of second order in equations.

In general coordinates, the variation of action $S[\Psi]$ under~\eqref{gs-aksz} is given by
\footnote{In deriving the expression for variation the following formula is useful
\begin{equation}
 \delta(\derham\Psi^A\psymp_A)=\delta\Psi^B \derham\Psi^A \sigma_{AB}=
 (-1)^{|B|(n-1)}\derham\Psi^A\sigma_{AB}\delta\Psi^B\;.
\end{equation}
}
\begin{equation}
\label{S-var}
 \delta_\lambda S[\Psi]=\int_X  \Big( -\lambda^C\d_C \cK -(-)^{|A|} \eom^A \lambda^C\d_C \cT_A+\half (-)^{|A|+|B|}\eom^B \eom^A \lambda^C \d_C \sigma_{AB}  \Big)\,,
\end{equation}
where
\begin{equation}
\label{LTT}
\eom^A= \derham \Psi^A + Q^A\,, \qquad {\cT}_A = \d_A \SM- Q^B\sigma_{BA} \,, \quad \cK =  Q^A \d_A \SM - \half Q^A Q^B \sigma_{BA}  \,.
\end{equation}
In the case of AKSZ model $\cK=0$ and $\cT_A=0$ so that the first two terms vanish identically, while the last term is
precisely quadratic in the equations of motion $\eom^A = 0$ \eqref{eom-aksz}. One concludes that modulo on-shell vanishing terms
transformation ~\eqref{gs-aksz}
is a gauge symmetry of action $S[\Psi]$. Let us note that in AKSZ  setting it is not difficult to
find an explicit expression for genuine gauge symmetries. However, in the generalization we consider next the
above indirect approach turns out to be useful. Note also that by construction AKSZ model is diffeomorphism
invariant. Moreover, the diffeomorphisms seen as gauge transformations  are just particular combinations of
gauge symmetries~\eqref{gs-aksz}.

\subsection{Relaxing AKSZ conditions}
\label{sec:relax}

Now we relax basic  axioms of the AKSZ model discussed above. We restrict ourselves to finite-dimensional $\manM$ and try
to find a generalization suitable for describing theories with local degrees of freedom. Recall that AKSZ with
finite-dimensional $\manM$ is necessarily topological in spacetime dimension $n>1$.

The main condition we give up here is the invertibility of the $2$-form $\sigma$. Furthermore, we do not immediately
assume that combinations $\cK$ and $\cT_A$ from \eqref{LTT} are vanishing.  In addition, we admit that not all
coordinates $\Psi^A$ with $\gh{\Psi^A}>0$ give rise to nontrivial gauge parameters.

The equations of motion following from~\eqref{masteraction} are given by
\begin{equation}
\eom_B\equiv \derham \Psi^A \sigma_{AB}+\d_B \SM=0\,.
\end{equation}
In the AKSZ case, these  are equivalent to $\eom^A=0$ that can be easily seen
by introducing inverse matrix  $(\sigma^{-1})^{AB}$.

The condition that transformations \eqref{gs-aksz} preserve action \eqref{masteraction}
modulo terms quadratic in the equations of motion takes the form
\begin{equation}
\label{29}
 \delta_\lambda S[\Psi]=\int_X \big((-)^{|A|+|B|} \eom_A \eom_B  \lambda^C R_C^{BA}+\text{total derivatives}\big)\,,
\end{equation}
where $R^{AB}_C$ are some local functions of fields.

Using~\eqref{S-var} and \eqref{29}  along with  the identity $\eom^A\sigma_{AB}=\eom_B - \cT_B$, and requiring the coefficients at
0th, 1st, and 2nd orders in $\eom^A$ to vanish one finds
\begin{equation}
\label{R-cond}
\ba{c}
\dps \lambda^C \d_C \cK + (-)^{|A|+|B|} \,\cT_A \cT_B \lambda^C R_C^{BA} = 0\;,
\\
\\
\dps
\half \lambda^C \d_C \cT_A +  (-)^{|A|+|B|+|D|+1}\sigma_{AD} \cT_B \lambda^C R_C^{BD}= 0\;,
\\
\\
\dps
\half \lambda^C \d_C \sigma_{NM}+
(-)^{|B| + |N|(|A|+|M|+n)+n+1}\sigma_{MA}\sigma_{NB}\lambda^C R_C^{BA}= 0\,.
\ea
\end{equation}
Obviously, the above expressions  are not covariant under general coordinate transformations on $\manM$
unless 2-form $\sigma$ nondegenerate. Note that even in this case a quantity $R^{AB}_C$ is not a tensor but
rather a connection. For the system to have a clear geometrical interpretation one therefore should
assume that $\manM$ is to be
equipped with additional structures and work in the special coordinate systems only. Leaving the study of associated
geometry and most general axioms for future work we now formulate some minimal (perhaps too restrictive) set of
axioms that guarantee that the gauge theory under consideration is consistent. We shall see  that even this partial
setting is sufficient for a variety of meaningful examples, see Section \bref{sec:examples}.

\subsubsection{Basic structures and compatibility conditions}

Without trying to be exhaustive we now assume $\manM$ to have a structure of a trivial vector bundle
$\manM=\manM_0\times \manM_1$ and refer to $\manM_0$ and $\manM_1$ as horizontal and vertical submanifolds,
respectively. By analogy with the AKSZ case, let $\manM$ be equipped with the following structures.

\begin{itemize}
\item Ghost degree $\gh{}$. For simplicity we assume that $\gh{\Psi^A}\geq 0$.
However, in general it can be useful to allow for negative degrees as in AKSZ or BV setting.

\item
1-form (pre-symplectic potential) $\chi=d\Psi^A\psymp_A $ such that $\gh{\psymp_A}=n-1-\gh{\Psi^A}$,
where $n$ is the positive integer.
\item
Odd vector field $\dps Q=Q^A(\Psi)\dl{\Psi^A}$ such that $\gh{Q}=1$ (note that $Q$ is not necessarily nilpotent).
\item
Function $\SM$ (potential or generalized  Hamiltonian), $\gh{\SM}=n$.

\end{itemize}

In addition, we introduce the differential subalgebra $I$ in the algebra of differential forms on $\manM$
determined by the decomposition $\manM_0\times \manM_1$. Let $\phi^\alpha, v^i$
be adapted coordinates on $\manM$, \textit{i.e.}, $\phi^\alpha$ are coordinates on $\manM_0$ and $v^i$ are
coordinates on the fibres $\manM_1$. Then, subalgebra $I$ is generated by $\phi^\alpha$, $\dm \phi^\alpha$ and $\dm v^i$.

Note that just like in the AKSZ setting the structures defined on $\manM$ depend on the positive integer $n$ to be identified with the space-time dimension. This is in contrast to
the formulation at the level of equations of motion where the target space structures are not aware of which space-time manifold is involved.

The compatibility conditions read as
\begin{align}
\sigma&\in I\,, \label{c1}\\
\cT=\dm \SM -i_Q \sigma&\in I\,,\label{c2}\\
 \cK=Q\SM-\half\sigma(Q,Q)&\in I\,,\label{c3}
\end{align}
where $\sigma(Q,Q)=i_Q i_Q \sigma$ and $\cT=d\Psi^A \cT_A$, cf. \eqref{LTT}.  These are precisely the conditions ~\eqref{R-cond} with $R^{AB}_C$ put to zero.
Note that $1$-form $\psymp$ enters the above expressions only through
$2$-form $\sigma=\dm \psymp$, while the second condition implies
$\Lie_Q\sigma\in I$, where $\Lie_Q$ denotes Lie derivative along $Q$. Indeed, acting by $d$ on both sides of
\eqref{c2} one obtains $d i_{Q} \sigma \in I$, and then uses formula \eqref{superlie}.
The second condition generalizes that of $Q$ being a Hamiltonian vector field with the Hamiltonian $\SM$.
The third condition is much less trivial and is a generalization of the master equation for $\SM$
or the nilpotency condition for $Q$.

Provided that  $\Psi^A$ are the  adapted coordinates $\phi^\alpha,v^i$ on $\manM$,
the component versions of  \eqref{c1} - \eqref{c3} read as
\begin{equation}
\label{compcond2}
\begin{gathered}
\dl{v^i} \sigma_{AB}=0\,, \qquad \dl{v^i} \cT_A= \dl{v^i}(\d_B \SM-Q^A\sigma_{AB})=0\,, \\
\dl{v^i}\cK=\dl{v^i}(Q^A\d_A \SM-\half Q^B  Q^A\sigma_{AB})=0\,.
\end{gathered}
\end{equation}


\subsubsection{Gauge symmetries}
\label{sec:Lagrangian}
By construction, action \eqref{masteraction}  is invariant under diffeomorphisms. In addition, consider the following transformations
\begin{equation}
 \delta_\lambda \Psi^A=\derham\lambda^A-\lambda^C\d_C Q^A\,,
\label{gs}
\end{equation}
where $\lambda^A$ is a gauge parameter associated to the field $\Psi^A$ as described below
\begin{equation}
\left.
\Psi^A=\begin{cases}
\phi^\alpha\qquad &         \lambda^\alpha=0 \\
v^i, \gh{v^i=0}\qquad & \lambda^i=0\\
v^i, \gh{v^i>0}\qquad & \lambda^i\neq 0
\end{cases}
\right\} = \lambda^A
\end{equation}
In other words, if $\Psi^A$ is a horizontal coordinate
$\phi^\alpha$, then $\lambda^\alpha=0$, while if $\Psi^A$ is a vertical coordinate  $v^i$, $\gh{v^i}=p>0$,
then $\lambda^i$ is non-zero, $\gh{\lambda^i} = p-1$. If $\gh{v^i}=0$, then an associated gauge
parameter is absent,  $\lambda^i=0$.

In terms of coordinates $\phi^\alpha, v^i$ transformations~\eqref{gs} take the form
\begin{equation}
\label{gaugeVH}
 \delta v^i=\derham\lambda^i-\lambda^j\d_j Q^i\,,\qquad \delta \phi^\alpha=-\lambda^j\d_j Q^\alpha\,,
\end{equation}
where $\d_i=\dl{v^i}$ and $\d_\alpha=\dl{\phi^\alpha}$. This explains the need to introduce the direct product
structure $\manM = \manM_0 \times \manM_1$: the vertical fields have associated differential
gauge parameters, while horizontal fields have not. Finally, with this choice of $\lambda$ conditions
\eqref{c1}-\eqref{c3} imply  \eqref{R-cond} (with vanishing $R^{AB}_C$) and the action~\eqref{masteraction} is gauge invariant.

Note that setting $R^{AB}_C =0$ is possible only if one uses
special coordinate systems (more precisely those induced by
coordinates on the factors $\manM_0$ and $\manM_1$). A formulation
in generic coordinates on $\manM$ would require an  introduction
of nontrivial $R^{AB}_C$ along with a certain covariant
differential on $\manM$. These interesting and important issues
will be considered elsewhere.

\subsubsection{The stronger compatibility conditions}
\label{sec:moreaxioms}

The compatibility conditions \eqref{c1}  - \eqref{c3} are invariant under
$\SM \to \SM+h$ with $h\in I$ such that $Qh\in I$. It could be convenient to decompose $\SM$
as $\SM^0-\HM$ where the ``minimal'' $\SM^0$ satisfies~\eqref{c2}, while $\HM$ is a generic ghost degree $n$
element from $I$ satisfying $Q\HM\in I$.

In particular, suppose one can choose $\psymp$ such that $\sigma = d\psymp$ satisfies \eqref{c1}  and
\begin{equation}
\label{Lchi}
\Lie_Q \psymp \in I\;.
\end{equation}
In this case using $L_Q=i_Q \dm -\dm i_Q$ one finds
$\dps\dm (i_Q \psymp)-i_Q\sigma\in I$ so that "minimal" $\SM^0$ chosen as $\SM^0=i_Q\chi$ satisfies~\eqref{c2}, where $i_Q\chi = Q^A\chi_A$.
The remaining freedom in $\SM$ is
then described by the potential $\HM$, so that a general $\SM$ satisfying \eqref{c2} can be represented as
\begin{equation}
\label{LQchi}
\SM = Q^A \chi_A - \HM\;,
\qquad
\HM\in I\;\;\text{and}\;\; Q\HM \in I\;.
\end{equation}


Let us consider now condition~\eqref{c3}. It is enough to check the condition for $\SM=\SM^0$. Condition \eqref{c3} can be then written as $i_Q\dm i_Q\psymp-\half i_Q i_Q \dm\psymp\in I$.
Using \eqref{inner}, \eqref{superlie}, \eqref{liechi}, along with the
identity $\dps i_Q\dm i_Q\psymp=\half \Lie_Q i_Q\psymp +\half i_Q\dm i_Q\psymp$,
one identically rewrites condition \eqref{c3} as
\begin{equation}
\commut{\Lie_Q}{i_Q}\chi \equiv  i_{Q^2}\psymp\in I\;.
\end{equation}
It follows that for nilpotent $Q$  condition  \eqref{c3} is identically satisfied.

To summarize: if one starts with nilpotent
$Q$ and $\chi,\HM$ such that $\Lie_Q\chi\in I$, $\HM\in I$, $Q\HM \in I$ then all the compatibility
conditions  are fulfilled.
In all the examples we consider below the basic objects satisfy this more restrictive conditions.

Quite often condition \eqref{Lchi} takes a strong form $\Lie_Q \psymp=0$ so that $\dm \SM=i_Q \sigma$.
Then,  condition~\eqref{c2} amounts to $Q\SM=0$ and $\sigma(Q,Q)=0$.

\section{Examples}
\label{sec:examples}

\subsection{The Einstein gravity}
\label{sec:gravity}

Let us consider the frame formulation of ordinary gravity. To apply the general construction developed in
Section \bref{sec:framelike} one takes $\manM=\Pi\algg$ with
Poincar\'e algebra $\algg = iso(n-1,1)$
\begin{equation}
\label{Poincare}
\begin{array}{c}
[L_{ab}, L_{cd}]  = \eta_{ac}L_{db}  - \eta_{bc}L_{da} - \eta_{ad}L_{cb} + \eta_{bd}L_{ca}\;,
\\
\\
\dps
[P_a, L_{bc}] = \eta_{ab}P_c  - \eta_{ac}P_b\;,
\qquad
[P_a, P_b] = 0\;,
\end{array}
\end{equation}
where $\eta_{ab}$ is a flat canonical Minkowski metric, and indices run $a,b,...=0,\ldots,n-1$.
Coordinates on $\manM$ are Grassmann odd
$e^a$ and $\omega^{ab} = -\omega^{ba}$ so that an arbitrary element in $\manM$ is parameterized
as $\Psi = e^a P_a + \half\, \omega^{ab} L_{ab}$.  Coordinates
$\omega^{ab}$ are vertical, while coordinates $e^a$ are horizontal. Odd vector field $Q$ on $\manM$ is given
by components
\begin{equation}
\label{Poincarediff}
 Q e^a=\omega^a{}_c\, e^c\,, \qquad Q\omega^{ab}=\omega^{a}{}_c \,\omega^{cb}\,,
\end{equation}
so that $Q$ is the Chevalley-Eilenberg differential for the Poincar\'e algebra, $Q^2 = 0$.

As a presymplectic potential $\chi$ we take
\begin{equation}
\label{presympot}
\chi = d e^a \chi_a + \half d \omega^{ab} \chi_{ab} = \frac{1}{(n-2)!}\epsilon_{abm_1\ldots m_{n-2}}
d \omega^{ab}\ee^{m_1}\ldots \ee^{m_{n-2}}\;.
\end{equation}
It is $Q$-invariant thanks to $o(n-1,1)$-invariance of the  Levi-Civita tensor $\epsilon_{a_1\ldots\, a_n}$ in $n$ dimensions.

The associated presymplectic $2$-form $\sigma$
computed according to \eqref{2formcomp} reads as
\comm{Some more explicit formulas
\begin{equation}
\begin{array}{l}
\dps
\sigma(e) = \frac{1}{2}\big[ d e^a d e^b \sigma_{a|b}+ d \omega^{ab} d e^c \sigma_{c|ab}
+ \frac{1}{4} d \omega^{ab} d \omega^{cd} \sigma_{ab|cd}\big] =
\\
\\
\dps
\hspace{45mm} = \frac{1}{(n-3)!}\, \epsilon_{abc\, m_1\ldots m_{n-3}} d \omega^{ab} d e^c
\ee^{m_1}\ldots \ee^{m_{n-3}}\;,
\end{array}
\end{equation}
}
\begin{equation}
\label{presympl2form}
 \sigma= d \omega^{ab} d e^c \Vol_{abc}\;.
\end{equation}
Here and in what follows we use the following notation for the generalized volume forms
\begin{equation}
\label{vol}
\Vol_{a_1 ... a_p} = \frac{1}{(n-p)!}\, \epsilon_{a_1 ... a_p c_1\ldots c_{n-p}} \ee^{c_1}\ldots \ee^{c_{n-p}}\;,
\qquad p = 0,..., n\;,
\end{equation}
that satisfy the identity
\begin{equation}
\label{VolIden}
e^c \Vol_{a_1 ... a_p} = \Vol_{[a_1 a_2 ... a_{p-1}} \delta^c_{a_p]} \;,
\end{equation}
where all  indices  $a_i$ are antisymmetrized with a unit weight.

Choosing potential $\SM$ as
\begin{equation}
\label{funcSgrav}
\SM=Q^A \chi_A = \omega^a{}_c \omega^{cb} \Vol_{ab}\,,
\end{equation}
the corresponding action functional \eqref{masteraction} takes the familiar form
\begin{equation}
\label{actgravity}
S_{GR}[e,\omega]=\int \big[\derham\omega^{ab}+\omega^{a}{}_c\omega^{cb}
\big] \,\Vol_{ab}\,.
\end{equation}
Here and in what follows the integral is taken over space-time manifold $X$ unless otherwise specified.
This is precisely the standard frame-like action of the Einstein gravity with zero cosmological constant. The expression in parenthesis is identified with the $2$-form Lorentz curvature
$R^{ab} = \derham\omega^{ab}+\omega^{a}{}_c\omega^{cb}$. Here,  the ghost number zero component fields $e^a=e^a_\mu d x^\mu,\omega^{ab}=\omega^{ab}_\mu d x^\mu$
enter $e^{a}(x,\theta)$ and $\omega^{ab}(x,\theta)$ as follows
\begin{equation}
 \ee^a=\theta^\mu e_\mu^a\,, \qquad \omega^{ab}=\theta^\mu \omega_\mu^{ab}\,.
\end{equation}

One can explicitly check that all the compatibility conditions \eqref{compcond2} are fulfilled.
Using \eqref{funcSgrav} one shows  that $L_Q \psymp=0$ because the Levi-Civita tensor is Lorentz invariant, while both
$e^a$ and $\derham\omega^{ab}$ are transformed by $Q$ as a vector and bivector, respectively. Following
the discussion of Section \bref{sec:moreaxioms} one then easily proves the compatibility conditions
in the form \eqref{c1} - \eqref{c3}.

Action \eqref{actgravity} can be non-trivially extended provided function $\SM$  \eqref{funcSgrav}
is augmented  as follows
\begin{equation}
\label{grav-SM}
\SM \rightarrow \SM - \HM\,,
\end{equation}
where $\HM $ is a new function such that $\HM \in I$, $Q\HM\in I$ and $\gh{\HM}=n$. Taking into account the ghost degree and the compatibility conditions the
only nontrivial choice is to take $\HM$ proportional to the volume
$n$-from $\frac{1}{n!}\Lambda\,\epsilon_{c_1\ldots c_{n}} \ee^{c_1}\ldots \ee^{c_n} \equiv \Lambda \Vol$, where we
generally trade prefactor $\Lambda$ for the cosmological constant. The action
with added term \eqref{grav-SM} is still invariant under local Lorentz symmetry transformations and therefore
vertical and horizontal fields as well as vector field $Q$ remain unchanged.
Another way to obtain the action with \eqref{grav-SM} is to consider $o(d-1,2)$-covariant formalism
for (anti-)de Sitter  gravity, see Section \bref{sec:MMSW}.

\vspace{-2mm}

\subsubsection{Gauge symmetries.}
Gauge transformations \eqref{gaugeVH} with parameter $\lambda^{ab}(x)$ are just   conventional local Lorentz gauge transformations
\begin{equation}
\delta^{Lor}_\lambda\, e_\mu^a =  - \lambda^{a}{}_b\, e^b_\mu \;,
\qquad
\delta^{Lor}_\lambda\, \omega_\mu{}^{ab} = \d_\mu \lambda^{ab} - \lambda^{a}{}_c\,\omega_{\mu}{}^{cb}
+ \lambda^{b}{}_c\,\omega_{\mu}{}^{ca}\;.
\end{equation}
The remaining gauge symmetries of the action \eqref{actgravity}
are precisely diffeomorphisms
\begin{equation}
\delta^{diff}_\xi e^a_\mu  = \xi^\nu \d_{\nu} e^a_\mu+ \d_\mu \xi^\nu e^a_\nu  \;,
\qquad
\delta^{diff}_\xi \omega^{ab}_\mu  =\xi^\nu \d_{\nu} \omega^{ab}_\mu + \d_\mu \xi^\nu \omega^{ab}_\nu  \;.
\end{equation}

The point is that in contrast to Lorentz transformations  frame-like gravity action \eqref{actgravity} is not invariant with respect to transformations
originating in translation subalgebra of the Poincar\'e algebra. The respective
variation is proportional to $\int R^aR^{bc} \lambda^d \epsilon_{abcd m_1 ... m_{d-4}} e^{m_1} ... e^{m_{d-4}}$,
where $\lambda^a$ is a gauge parameter associated with translations, \textit{i.e.}, with the  frame field $e^a$, while $R^{ab}$ and $R^a$ are Lorentz curvature and the torsion. Obviously, in the linearized gravity the symmetry is restored.

At the nonlinear level one can still relate Poincar\'e translations
to diffeomorphisms through
\begin{equation}
\delta^{transl}_\zeta e_\mu^a = \delta^{diff}_\xi e_\mu^a
+ \delta^{Lor}_\lambda e_\mu^a + R_{\mu\nu}^a\xi^\nu\;,
\end{equation}
where diffeomorphism $\xi_\mu$ and  frame $\zeta^a$ vector parameters are related as $\xi_\mu = e^a_\mu \zeta_a$,
and $\lambda^{ab} = \omega^{ab}_\nu \xi^\nu$. Moreover, on the stationary surface diffeomorphisms can be expressed just in terms of translations and Lorentz rotations.
Indeed, the equation of motion $\delta S_{GR} /\delta \omega_\mu^{ab}$ = 0 implies that the torsion vanishes, $R_{\mu\nu}^a = 0$ so that
on-shell one can identify Poincar\'e translations with diffeomorphisms. Furthermore, the vanishing torsion constraint
expresses Lorentz connections via derivatives of the frame fields.
It turns out that for the gravity the vertical and horizontal fields are respectively
auxiliary and dynamical ones.

One concludes that the splitting between horizontal and vertical coordinates in the target space $\manM$ naturally fits the gauge structure of the Poincar\'e gauge gravity. Vertical gauge transformations along with diffeomorphisms produce translational symmetry not seen  within the geometrical setting of the theory.

\subsection{Gravity + scalar field}
\label{sec:gravity+scalar}
To describe scalar field coupled to gravity we extend the superspace  $\manM$ of Section \bref{sec:gravity} by the horizontal coordinates $\phi$,
$\gh{\phi}=0$ and $\pi^a$, $\gh{\pi^a}=0$. Additional components of the presymplectic potential \eqref{presympot}
and odd vector field \eqref{Poincarediff} are
\begin{equation}
\psymp(\dl{\pi^a})=0\,,
\qquad
\psymp(\dl{\phi})= \pi^a \Vol_a\,,
\end{equation}
and
\begin{equation}
Q\phi=0\,, \qquad Q\pi^a=\omega^a_b \pi^b\;.
\end{equation}
Note that $Q^2 = 0$. By analogy with the pure gravity, condition $L_Q\psymp=0$ immediately follows from the
invariance of the Levi-Civita tensor.

Consider function  $\SM=Q^A \psymp_A-\HM$. As $\HM$ we take scalar field covariant Hamiltonian
(in the sense of de Donder--Weyl formalism)
multiplied by the volume form, \textit{i.e.},
\begin{equation}
\HM=(-\frac{1}{2}\pi^a\pi_a+\frac{1}{2} m^2\phi^2)\,\Vol\;,
\end{equation}
which is $Q$-invariant and depends on horizontal coordinates only. Coupling $m^2$ is the mass of a scalar.
The entire action is familiar and reads as
\begin{equation}
\label{grav+scalar}
S[e,\omega,\phi, \pi]=S_{GR}[e,\omega]+\int \big[\,\derham \phi\, \pi^{a} \Vol_{a} +
\half (\pi^a\pi_a-m^2\phi^2)\,\Vol\,\big]\;,
\end{equation}
where the gravity action $S_{GR}[e,\omega]$ is given by \eqref{actgravity}.
Using the identity \eqref{VolIden} and eliminating $\pi^a$ one obtains
\begin{equation}
S[e,\omega,\phi]=S_{GR}[e, \omega]-\half \int
(g^{\mu\nu}\d_\mu \phi\, \d_\nu \phi+m^2\phi^2)\Vol\,, \qquad g^{\mu\nu}:=e_a^\mu  e^{a\nu}\;.
\end{equation}

It is also instructive to write action~\eqref{grav+scalar} over the Minkowski background described
by connections $\omega^{ab}=0$ and $e^a=\derham x^a$. Then,  action  \eqref{grav+scalar} takes the standard form
\begin{equation}
\label{scalar-1st}
S[\phi,\pi] =\int d^n x (\pi^a\d_a\phi+\half \pi^a\pi_a-\half m^2\phi^2)\;,
 \end{equation}
which is the well-known $1$st order massive scalar field action.
In parallel, in the flat case the scalar field contribution to the presymplectic 1-form $\chi$ becomes
\begin{equation}
\frac{1}{(n-1)!} \,
d \phi\, \pi^a dx^{b_1}\ldots dx^{b_{n-1}} \epsilon_{ab_1\ldots b_{n-1}}\;.
\end{equation}
It is a version of the canonical $n$-form of the covariant Hamiltonian
formulation for the scalar field, see, \textit{e.g.}, ~\cite{Gotay:1997eg}.

\subsection{Einstein-Yang-Mills theory}
\label{sec:gravity+YM}
Consider Yang-Mills theory with the gauge algebra $\algg$, which is given by basis elements $[T^I, T^J]  = C^{IJ}_K T^K$
and have the Killing form $ \left< T^I, T^J\right> = \delta^{IJ}$.
To describe Yang-Mills theory minimally coupled to gravity we extend the superspace $\manM$ of the
pure gravity of Section \bref{sec:gravity} with the following extra vertical coordinate $A^I$, $\gh{A}=1$
and extra horizontal coordinate $F^I_{ab} = - F^I_{ba}$, $\gh{F^I_{ab}}=0$.

The new components of the presymplectic potential \eqref{presympot} and odd vector field
\eqref{Poincarediff} are
\begin{equation}
\psymp(\dl{A^I})=F_I^{ab}\, \Vol_{ab}\;,
\qquad
\psymp(\dl{F^I_{ab}})=0\;,
\end{equation}
and
\begin{equation}
QA=\half\commut{A}{A}\,,\qquad QF_I^{ab}=\omega^a{}_c F_I^{cb}+\omega^b{}_c F_I^{ac}+\commut{A}F^{ab}_I\,.
\end{equation}
One can identify $Q$ as a Lie algebra cohomology differential for a direct sum of the Poincar\'e and
the YM gauge algebra represented on $F^{ab}_I$. That $L_Q\psymp=0$ can be checked as follows. The YM
contribution to $\psymp$ can be written as
\begin{equation}
 \psymp_{YM}=\left< dA ,\, F^{ab}\right>\,\Vol_{ab}\;.
\end{equation}
Decomposing $Q$ as $Q_{GR}+Q_{YM}$ one finds that $L_{Q_{GR}}\psymp_{YM}=0$ in exactly the same way as before,
while $L_{Q_{YM}}\psymp_{YM}=0$ follows from the invariance of the Killing form  and the relation $L_{Q_{YM}} dA=\commut{A}{dA}$.

Function $\SM=\psymp_A Q^A-\HM$ is given explicitly by
\begin{equation}
\SM=L_{GR}+\left<F^{ab},\commut{A}{A}\right> \Vol_{ab}
-\HM\;.
\end{equation}
As $\HM$ we again take the de Donder--Weyl generalized Hamiltonian of the YM field multiplied by the
volume form
\begin{equation}
\HM=- \Vol\,\left< F^{ab}, F_{ab} \right>\;,
\end{equation}
which again depends on horizontal coordinates only and  satisfies $Q\HM=0$.

The action takes the familiar form
\begin{equation}
\label{grav+YM}
S[e,\omega,A,F]=S_{GR}[e,\omega]+
\int  \left< \derham A+\half\commut{A}{A}\,,\, F^{ab}\right> \Vol_{ab}
+ \int \left<F^{ab}, F_{ab}\right>\Vol\;,
\end{equation}
where the gravity action $S_{GR}[e,\omega]$ is given by \eqref{actgravity}.
Using the identity \eqref{VolIden} and eliminating $F^{ab}_I$ one obtains
\begin{equation}
S[e,\omega,A]=S_{GR}[e,\omega]-\frac{1}{4}
\int g^{\mu\rho}g^{\nu\sigma}\langle\cF_{\mu\nu}, \cF_{\rho\sigma}\rangle\,,
\end{equation}
where $\cF^I_{\mu\nu}:=\d_\mu A^I_\nu-\d_\nu A^I_\mu +[A_\mu,A_\nu]^I$.

As before, in the flat background action~\eqref{grav+YM} reduces to the well-known $1$st order Yang-Mills  action
(actions of this type were originally proposed in~\cite{Arnowitt:1962hi})
\begin{equation}
S[A] =  \int d^nx(F_I^{ab}(\d_a A^I_b-\d_b A^I_a+\commut{A_a}{A_b}^I)-F^I_{ab}F_I^{ab})\;.
\end{equation}

\subsection{Multi-frame theory}
\label{sec:multiframe}

One can extend a target $\manM$ of Poincar\'e gravity  by adding $\cN$ copies of $\Pi \algg$ with $\algg$  Poincar\'e algebra~\eqref{Poincare} so that
$\manM = \Pi (\algg \oplus  \algg \oplus ...  \algg)$. Grassmann odd variables
$\omega^{ab}(i)$ and $e^a(i)$ where $i = 1, ..., \cN$, are vertical and horizontal, respectively. Odd vector
field generalizes \eqref{Poincarediff} as
\begin{equation}
\label{PoincarediffMASS}
Q e^a(i)=\omega^a{}_c(i)\, e^c(i)\,, \qquad Q\omega^{ab}(i)=\omega^{a}{}_c(i) \,\omega^{cb}(i)\,.
\end{equation}
The presymplectic forms on $\manM$ are then sums of those on $\Pi \algg$, cf. \eqref{presympot}, \eqref{presympl2form},
\begin{equation}
\label{premass}
\chi  = \sum_{i=1}^\cN \chi(i)\;,
\qquad
\sigma = \sum_{i=1}^\cN \sigma(i)\;.
\end{equation}
As $\SM$ we chose a direct generalization of that in the case of gravity:
\begin{equation}
L = \sum_{i = 1}^\cN
Q^A \chi_A(i)  +
\sum_{i=1}^{\cN}\frac{1}{n!}\,\Lambda(i)\,\epsilon_{m_1\ldots m_{n}} \ee^{m_1}(i)\ldots \ee^{m_n}(i)
\end{equation}
where $\Lambda(i)$ are the cosmological constants associated to each gravity sector.

To introduce multi-graviton interactions one mixes  frames from different sectors adding terms
like $\epsilon_{a_1 ... a_n}\,
e^{a_1}(i_1) \cdots e^{a_n}(i_n)$ \cite{Hinterbichler:2012cn}. However,
such terms break $\cN$  local Lorentz symmetry groups \eqref{PoincarediffMASS} down to the diagonal subgroup.
To restore  $\cN-1$ Lorentz symmetries one introduce $\cN-1$ compensating fields.

To this end one further extends $\manM_\cN$ to include   new Grassmann odd coordinates  $K^a{}_b(j)$, $j = 2,..., \cN$
satisfying the matrix constraint $K^{T} \eta K = \eta$, so that the presymplectic $1$-from \eqref{premass}
is unchanged, while odd vector field \eqref{PoincarediffMASS} acquires new
components
\begin{equation}
\label{neqQstueck}
Q K^{a}{}_b(j) =  \omega^{a}{}_c(1) K^{c}{}_b(j) -K^{a}{}_c(j) \omega^{c}{}_b(j)\;,
\qquad j = 2,..., \cN \;.
\end{equation}
Fields $K^a{}_b$ are horizontal. These are introduced in such a way that $Q$ acts
on  frames  $\bar e^b(j) \equiv K^a{}_b(j) e^b(j)$ as follows
\begin{equation}
Q[K^a{}_b(j) e^b(j)] = \omega^a{}_b(1) [K^b{}_c(j) e^c(j)]\;,
\end{equation}
so that $\bar e^a$ can be used to build interaction cross-term supporting an overall local Lorentz
invariance.

Leaving intact the presymplectic forms on $\manM$ one modifies function $\SM$
as follows
\begin{equation}
\label{multiL}
L =
Q^A \chi_A
+\sum_{i_1, ... , i_n = 1}^{\cN}\beta^{i_1 ... i_n}\, \epsilon_{a_1 ... a_n}\,
\big[K^{a_1}{}_{b_1}(i_1) e^{b_1}(i_1)\big] \cdots \big[K^{a_n}{}_{b_n}(i_n) e^{b_n}(i_n)\big]\;,
\end{equation}
where $\beta^{i_1 ... i_n}$ are totally symmetric coupling constants, while $K^a{}_b(1) = \delta^a_b$.
Then, the multi-frame  action takes the form\footnote{This actions is similar to that proposed in
\cite{Hassan:2012wt}, where the authors introduced $\cN$ St\"ueckelberg fields and operate with $\cN+1$
unbroken local Lorentz invariances. Action \eqref{multaction} contains the  minimum required number of $\cN-1$
St\"ueckelberg fields needed to compensate $\cN-1$ broken local Lorentz symmetries.  }
\begin{multline}
\label{multaction}
S[e,\omega, K] = \int \Big[ \sum_{i=1}^\cN  \big(\derham\omega^{ab}(i)+\omega^{a}{}_c(i)\omega^{cb}(i)
\big) \,\Vol_{ab}(i)
\\
+\sum_{i_1, ... , i_n = 1}^{\cN}\beta^{i_1 ... i_n}\, \epsilon_{a_1 ... a_n}\,
\big[K^{a_1}{}_{b_1}(i_1) e^{b_1}(i_1)\big] \cdots \big[K^{a_n}{}_{b_n}(i_n) e^{b_n}(i_n)\big]\Big]\;.
 \end{multline}
By construction, the above action is invariant with respect to local Lorentz symmetry transformations
read off from \eqref{PoincarediffMASS} and \eqref{neqQstueck},
\begin{equation}
\ba{c}
\delta^{Lor}_\lambda \omega^{ab}(i) = \derham \lambda^{ab}(i)  - \lambda^{ac}(i) \omega_c{}^b(i) + \lambda^{bc}(i) \omega_c{}^a(i)\;,
\\
\\
\delta^{Lor}_\lambda e^{a}(i) = - \lambda^{a}{}_b(i) e^b(i)\;,
\qquad
\delta^{Lor}_\lambda K^a{}_b(j) =  - \lambda^a{}_c(1) K^c{}_b(j) + \lambda^c{}_b(j) K^a{}_c\;,
\ea
\end{equation}
where $i = 1,..., \cN$ and $j = 2,..., \cN$. Fields $K^a{}_b(j)$ are St\"ueckelberg fields that compensate
broken local Lorentz symmetry. Using the gauge $K^a{}_b(j) = \delta^a_b$ one finds out that parameters
$\lambda^{ab}(i)$ are set to satisfy matrix constraints
\begin{equation}
- \lambda(1) K(j) + K(j)\lambda(j) = 0\;, \qquad j = 2,..., \cN\;,
\end{equation}
for the gauge fixed $K(j) = \mathbb{I}_n$.
It follows that $\lambda \equiv \lambda(1) = \lambda(2) = ... = \lambda(\cN)$, and the action retains  a
single  local Lorentz symmetry with a common parameter $\lambda$ for all frame fields.
The gauge fixed form of action \eqref{multaction} has been shown to describe  consistent interactions
of a single massless spin-$2$ field  and $\cN-1$ massive spin-$2$ fields \cite{Hinterbichler:2012cn,Hassan:2012wt}.

\subsection{First-order form of MMSW action for $AdS_n$ gravity}
\label{sec:MMSW}

\subsubsection{$n=4$}
Consider a superspace  $\manM_0$  with Grassmann odd, vertical coordinates $\omega^{AB}=-\omega^{BA}$, $\gh{\omega^{AB}}=1$,
and Grassmann even, horizontal
coordinates $V^A$ , $\gh{V^A}=0$ and $F^{AB}=-F^{BA}$, $\gh{F^{AB}}=2$.
Take as  $\manM \subset \manM_0$ the surface singled out by the following constraints
\begin{equation}
\label{MMSW-const}
 V^A V_A=-1\qquad F^{AB}V_B=0\,.
\end{equation}
Here and below indices $A, B, ... = 0,...,n$ are raised and lowered by a flat canonical Minkowski
metric $\eta_{AB} = (--+ ... +)$.
Odd vector field $Q$ with components
\begin{equation}
 Q\omega^{AB}=\omega^A{}_C \omega^{CB}\,, \qquad QV^A=\omega^A{}_B V^B\,,\qquad Q F^{AB}=\omega^A{}_C F^{CB}
 +\omega^B{}_C F^{AC}
\end{equation}
can be identified with the Chevalley-Eilenberg differential of $o(n-1,2)$ algebra  with coefficients in representation on
vector $V$ and tensor $F$. Odd vector field $Q$ is tangent to $\manM$ (\textit{i.e.}, constraints \eqref{MMSW-const}
are invariant on the constraint surface).

Specializing to $n=4$ we take a presymplectic potential $\psymp$  in the form
\begin{equation}
 \chi(\dl{V^A})=\chi(\dl{F^{AB}})=0\,,\qquad \chi(\dl{\omega^{AB}})=\epsilon_{ABCDE}F^{CD}V^E\,,
\end{equation}
while function $\SM$ is given by $\SM= Q^A\chi_A -\HM$, where
\begin{equation}
 \HM=-\half\epsilon_{ABCDE}F^{AB}F^{CD}V^E\,.
\end{equation}
One can show that all the compatibility conditions \eqref{c1}-\eqref{c3} are satisfied
due to $Q^2=0$ and the invariance of the $o(n-1,2)$ Levi-Civita tensor.

According to \eqref{masteraction} we build the following action
\begin{equation}
 S[\omega,V,F]=\int \epsilon_{ABCDE}(\derham\omega+\omega\omega)^{AB}F^{CD}V^E+\half\epsilon_{ABCDE}F^{AB}F^{CD}V^E\;.
\end{equation}
The above action can be shown to be dynamically equivalent to the standard MacDowell-Mansouri-Stelle-West
 (MMSW) action \cite{MacDowell:1977jt,Stelle:1979aj}, see formula \eqref{MMSW} below.
To this end, let us consider the Euler-Lagrange equation for $F^{AB}$ given by
\begin{equation}
\label{algeq}
 \epsilon_{ABCDE}F^{CD}V^E+\epsilon_{ABCDE}(\derham\omega+\omega\omega)^{CD}V^E=0\,.
\end{equation}
Since this equation is algebraic with respect to $F^{AB}$, it is enough to solve it at a given point.
By using the first equation in~\eqref{MMSW-const} one takes field $V^A$ in the form
$V^A=\delta^A_{(n)}$, and finds that $F^{ab}=-(\derham\omega+\omega\omega)^{ab}$ where $a,b=1,\ldots,n-1$.
Other components of $F$ vanish by virtue of  the second condition in~\eqref{MMSW-const}, and, moreover,
they do not contribute to either of the terms in the action. One then concludes that $F^{AB}$ are auxiliary fields,
and their elimination gives the MMSW gravity action
\begin{equation}
\label{MMSW}
S[\omega,V]=-\half\int \epsilon_{ABCDE} R^{AB}R^{CD}V^E\,,
\qquad
R^{AB}:= (\derham\omega+\omega\omega)^{AB}\;.
\end{equation}

\subsubsection{$n>4$}
\label{MMSV-g}

In addition to the variables introduced in the case $n=4$, let us also introduce
extra horizontal variables $E^A$, $\gh{E^A}=1$ and $\pi_A$, $\gh{\pi_A}=n-1$ subjected to
constraints
\begin{equation}
\label{consmmsw}
E^AV_A=0\;,
\qquad
\pi_AV^A=0\;.
\end{equation}
The nonvanishing components of presymplectic potential $\psymp$ are
\begin{equation}
\psymp(\dl{V^A})
=\pi_A\,,\quad
\psymp(\dl{\omega^{AB}})=\frac{1}{(n-2)!}\epsilon_{ABCD B_1\ldots B_{n-4} E}F^{CD}E^{B_1}\ldots E^{B_{n-4}}V^E\,.
\end{equation}
The odd vector field $Q$ determined by
\begin{equation}
\begin{gathered}
Q\omega^{AB}=\omega^A{}_C \omega^{CB}\,, \qquad Q F^{AB}=\omega^A{}_C F^{CB}+\omega^B{}_C F^{AC}\,,\\
QV^A=\omega^A{}_B V^B\,,\qquad Q E^A=\omega^A{}_B E^B\,,\qquad Q\pi_A=-\omega_A{}^B\pi_B\;,
\end{gathered}
\end{equation}
is identified with the Chevalley-Eilenberg differential of $o(n-1,2)$ algebra  with coefficients in representation on
vectors $V$, $E$, $\pi$, and tensor $F$. One can check that the constraints \eqref{consmmsw} are $Q$-invariant while $L_Q \psymp\in I$. More precisely,
\begin{equation}
 L_Q \psymp=
-d\omega^A{}_B V^B\pi_A\,.
\end{equation}
Note that in contrast to all the previous examples $L_Q\sigma\neq 0$.

Function $\SM$ is given by $\SM= Q^A\chi_A -\HM$, where
\begin{equation}
\HM=-\frac{1}{2(n-2)!}\epsilon_{ABCDC_1\ldots C_{n-4}E}F^{AB}F^{CD}E^{C_1}\ldots E^{C_{n-4}}V^E
- E^A \pi_A\;,
\end{equation}
so that the action takes the following form
\begin{multline}
 S[\omega,V,F,E, \pi]=\int \Big[
(\derham V^A+\omega^A{}_B V^B-E^A)\pi_A
+\\
\frac{1}{(n-2)!}\epsilon_{ABCDC_1\ldots C_{n-4}E}((d\omega+\omega\omega)^{AB}F^{CD}+\half F^{AB}F^{CD}) E^{C_1}\ldots E^{C_{n-4}} V^E
\big]\;.
\end{multline}
Fields $E_A$ and $\pi_A$ are clearly auxiliary. As before, the same is also true for $F^{AB}$.
Eliminating the auxiliary fields  by their own equations of motion one finally gets the standard action \cite{Vasiliev:2001wa}
\begin{equation}
S[\omega,V]=-\frac{1}{2(n-2)!}\int\epsilon_{ABCDC_1\ldots C_{n-4}E} R^{AB}R^{CD} E^{C_1}\ldots E^{C_{n-4}} V^E \,,
\end{equation}
where $E^A:=(dV+\omega V)^A$ and $R^{AB}:= (\derham\omega+\omega\omega)^{AB}$.


\subsection{Linearized frame-like actions}
\label{sec:linearized}

It is instructive to examine the perturbation theory for action \eqref{masteraction} over
background field values $\Psi_0^A$. The
fluctuations over the background are defined as $\Psi^A = \Psi_0^A+ \Phi^A$.
Then, one finds that modulo additive constants and total derivative terms the quadratic action reads
\begin{multline}
\label{masteractionPERT}
S_0[\Phi]=\int \big[\derham \Phi^A \Phi^B  \d_B\psymp_{A}(\Psi_0)
\\
+ \half\Phi^A \Phi^B \d_B\d_A \SM(\Psi_0) + \half\derham \Psi_0^C \Phi^A \Phi^B \d_B \d_A \psymp_{C}(\Psi_0)\big]\,.
\end{multline}
The above action is again of the form \eqref{masteraction}. Adding total derivatives all the dependence on $\chi_A$
can be expressed in terms of $\sigma_{AB}$. More precisely, adding
\begin{multline}
-\half\derham (\Phi^A \Phi^B \d_B\chi_A)=-\half \big[\derham \Phi^A \Phi^B(\d_B \chi_A(\Psi_0)+(-)^{|A||B|}\d_A\chi_B(\Psi_0))\\+(-1)^{|A|+|B|}
\Phi^A\Phi^B\derham\Psi^C_0\d_C\d_B\chi_A(\Psi_0)\big]
\end{multline}
to~\eqref{masteractionPERT} gives\comm{A useful trick in the derivations is to represent de Rham differential as $\derham=\derham \Phi^A \dl{\Phi^A} +\derham \Psi^A_0\dl{\Psi^A_0}$}
\begin{multline}
\label{pertval}
 S_0[\Phi]=-\half\int \big[(-)^{|A|}\derham \Phi^A \Phi^B \sigma_{BA}(\Psi_0)~-\\ -~\Phi^A \Phi^B (\d_B\d_A \SM(\Psi_0)-(-)^{|C|}\derham \Psi_0^C \d_B\sigma_{AC}(\Psi_0))\big]\,.
\end{multline}

The presymplectic $2$-form $\bar\sigma_{AB}=\sigma_{AB}(\Psi_0)$ entering the expression for the linearized action does not depend on field variables.  It may imply an extension of the compatibility conditions \eqref{compcond2} in such a way that the quadratic approximation
has more vertical fields than the original theory. Also, in deriving the quadratic action we do not assume  the original action
invariant under gauge symmetry transformations \eqref{gs} or \eqref{gaugeVH}.

%

\subsubsection{Linearized gravity}

Let us consider the gravity action \eqref{actgravity} linearized around Minkowski spacetime $\mathbb{R}^{n-1,1}$ given by
$\omega_0^{ab} = 0$ and $e_0^a = \derham x^a$. Let $e^a$ and $\omega^{ab}$ again denote dynamical fields.
The linearized presymplectic $2$-form is given by
\begin{equation}
\ba{c}
\dps
\bar\chi=  d \omega^{ab}e^c \bar\Vol_{abc}\;,
\qquad
\bar\sigma = d\omega^{ab} de^c \bar\Vol_{abc}\;,
\end{array}
\end{equation}
where the basis forms $\bar\Vol = \bar\Vol(e_0)$ \eqref{vol} are built of the background frame field $e_0^a$, while the linearized
Poincar\'e algebra differential is determined by
\begin{equation}
\label{PoincarediffPERT}
\bar Q e^a=\omega^a{}_c\, e_0^c\,, \qquad \bar Q \omega^{ab}= 0 \,.
\end{equation}
The linearized potential takes the form
\begin{equation}
\label{SM}
\SM  =  \omega^{a}{}_c\,\omega^{cb}\, \bar\Vol_{ab}\;.
\end{equation}
Finally, the linearized action \eqref{masteractionPERT}
takes the form
\begin{equation}
\label{actlingrav}
S_0[e,\omega] = \int
\Big[\derham\omega^{ab}e^c \bar\Vol_{abc} + \omega^{a}{}_c\omega^{cb} \bar\Vol_{ab} \Big]\;.
\end{equation}
Adding a total derivative and using \eqref{pertval} along with $\derham e_0=0$ satisfied by background
it can be equivalently represented as
\begin{equation}
\label{gravlinHS}
S_0[e,\omega] = \int
\Big[ \derham e^a  - \half \omega^{a}{}_{d} e_0^d \Big] \omega^{bc} \bar\Vol_{abc}\;,
\end{equation}
which  is manifestly invariant with respect to the gauge symmetry transformations $\delta_\lambda e^a = \derham \lambda^a$ and
$\delta_\lambda\omega^{ab} = 0$, originating from local translations. One concludes that all fields  can be treated now as vertical ones.
Indeed, the extended compatibility conditions  \eqref{compcond2} are valid in this case because
components of the presymplectic 2-form are field-independent

Action \eqref{gravlinHS} is directly generalized to higher spin fields which we consider next.

\subsubsection{Frame-like action for spin-$s$ massless fields}
\label{sec:HS}

Consider a target superspace  $\manM$ with odd coordinates
\begin{equation}
\label{Ms}
\omega^{a_1 ... a_{s-1},\, b_1 ... b_t}\;,
\qquad
t = 0, ..., s-1\;,
\end{equation}
which are irreducible Lorentz tensors satisfying tracelessness and Young symmetry conditions. Spin parameter $s$ is some integer
$s = 1,2,... $. We assume that all coordinates
are vertical ones. The odd vector field $Q$ on $\manM$ is given by components
\begin{equation}
 Q \omega^{a_1 ... a_{s-1},\, b_1 ... b_t}=\omega^{a_1 ... a_{s-1},\, b_1 ... b_t}{}_c \, e_0^c\;,
\end{equation}
where just like in the case of spin-2 field $e_0$ is an extra odd variable of ghost degree $1$ interpreted as a target space
parameter giving rise to background frame field. It is nilpotent, $Q^2 = 0$, and can be seen as a generalization
of the linearized spin-$2$ differential \eqref{PoincarediffPERT}.

In terms of field variables odd vector field $Q$ induces the following gauge transformations
\begin{equation}
\label{gaugeparameters}
\delta_\lambda \omega^{a_1 ... a_{s-1},\, b_1 ... b_t} = \derham \lambda^{a_1 ... a_{s-1},\, b_1 ... b_t} +
\lambda^{a_1 ... a_{s-1},\, b_1 ... b_t}{}_c e_0^{c}\;,
\end{equation}
where $e_0^c$ is the background frame field satisfying $\derham e^a_0=0$ (more generally, to work in terms of
generic frame one also introduces background Lorentz connection $\omega_0^{ab}$ so that $\derham e_0+\omega_0e_0=0$
and $\derham \omega_0+\omega_0\omega_0=0$, \textit{i.e.} the zero-curvature equations
of the Poincar\'e algebra). Fields $\omega^{a_1 ... a_{s-1},\, b_1 ... b_t}
=\derham x^\mu \omega_\mu^{a_1 ... a_{s-1},\, b_1 ... b_t}$ are differential $1$-forms on the spacetime manifold,
while the gauge parameters $\lambda^{a_1 ... a_{s-1},\, b_1 ... b_t}$ are $0$-forms.

Consider the following presymplectic $1$-form and potential
\begin{equation}
\begin{gathered}
\label{potsig}
\chi
=  d \omega^{am_1...m_{s-2}}\omega_{m_1...m_{s-2}}{}^{b,\,c} \bar\Vol_{abc}\;,
\\
%
\SM = \half  \omega^{am_1...m_{s-2},k}\omega_{m_1...m_{s-2}}{}^{b,\,c}e_0{}_k  \bar\Vol_{abc}\;.
\end{gathered}
\end{equation}
The associated 2-form reads as
\begin{equation}
\sigma
= d \omega^{am_1...m_{s-2}}d  \omega_{m_1...m_{s-2}}{}^{b,\,c} \,\bar\Vol_{abc}\;.
\end{equation}
The presymplectic $2$-form depends on the background frame field
only that conforms with the compatibility condition \eqref{c1}.

The above prerequisites are used to build an action functional  according to \eqref{masteraction} as follows
\begin{equation}
\label{linHS}
S_0[\omega] = \int \Big[\derham \omega^{am_1...m_{s-2}} -\half  \omega^{am_1...m_{s-2},k}e_0{}_k\Big] \omega_{m_1...m_{s-2}}{}^{b,\,c} \,\bar\Vol_{abc}\;,
\end{equation}
which is the frame-like action of massless spin-$s$ fields on Minkowski spacetime
\cite{Vasiliev:1980as}. This is a  generalization of  spin-$2$ case \eqref{gravlinHS}.

It is important to note that \eqref{linHS} does not depend on
fields $\omega^{a_1 ... a_{s-1},\, b_1 ... b_t}$ with $t \geq 2$
(called extra fields). This is a new feature not present in other
examples: the system contains fields that do not enter the action
and hence should not be considered dynamical. We do not discuss
here how this subtlety can be naturally handled in the formalism.
\footnote{The extra fields do contribute to the higher spin action
when considering interactions of FV-type
\cite{Fradkin:1986qy,Vasiliev:2001wa,Alkalaev:2002rq,Vasilev:2011xf,Boulanger:2012dx}.}
In the next section we propose an alternative formulation where
all the fields are at the equal footing. Furthermore,
in Section \bref{sec:onedim} we describe generic mechanical systems from the frame-like perspective that naturally leads
to the notion of extra fields as fields not entering a Lagrangian of a system but producing (a part of) its gauge symmetry invariance, see our comments below formula
\eqref{final-red}.

It is claimed that the above action is the most general one
containing only dimensionless  coefficients. Generally, one may
consider an action functional \eqref{masteraction} on $\manM$
\eqref{Ms} where all fields contribute and not only the two lowest
rank ones identified with the generalized frame and Lorentz
connections.  However, using the dimensional analysis one finds
out that other possible terms with extra fields in the
searched-for action necessarily contain dimensionful coefficients.
The extra field terms can either be set to zero directly
(\textit{i.e.}, the corresponding coefficients), or be combined
into total derivatives. Whence, the only scale invariant
combination is given by \eqref{linHS}. The analogous reasoning
applies in the case of $AdS$ background, where the higher spin
action of the type \eqref{masteraction} may have terms involving extra fields
with overall dimensional  coefficients proportional to inverse
powers of the cosmological constant. One may require  all such
terms to combine into total derivatives giving the
$AdS$ higher spin action of the form \eqref{linHS} with function
$\SM$ shifted as in \eqref{grav-SM}. This is in fact the extra
field decoupling condition of Lopatin and Vasiliev
\cite{Lopatin:1988hz}. \footnote{Higher spin Lagrangians of
\cite{Lopatin:1988hz} are  built as bilinear
combinations of linearized gauge invariant curvatures so that
these are manifestly gauge invariant. Modulo total derivative
terms these can be shown to be of the form \eqref{masteraction}.}

The above consideration of free higher-spin Lagrangians visualized
as the presymplectic AKSZ-type  models has many features in common
with the search for Lagrangians within the unfolded formulation
\cite{Vasiliev:2005zu,Vasilev:2011xf}. In particular, it can be
extended to other higher-spin  systems like
\cite{Skvortsov:2008sh,Zinoviev:2008ve,Ponomarev:2010st,Metsaev:2011iz}  as well as
to lower-spin  supersymmetric models given within the unfolded
formulation \cite{Misuna:2013ww}. Also, it would be interesting to
reconsider from this perspective the gauge symmetry structure of
the frame-like Lagrangians for $AdS$ mixed-symmetry massless
fields proposed in \cite{Alkalaev:2003qv,Alkalaev:2003hc,Alkalaev:2005kw}.

\subsubsection{Extended frame-like action for spin-$s$ massless fields}

The extra fields (\textit{i.e.}, $\omega^{a_1 ... a_{s-1},\, b_1 ... b_t}$
with $t \geq 2$) do not enter the action  \eqref{linHS}. However,
the gauge parameter $\lambda^{{a_1 ... a_{s-1},\, b_1 b_2}}$
associated to $\omega^{a_1 ... a_{s-1},\, b_1 b_2}$  is still
needed to describe gauge invariance of the action. More precisely, the gauge transformations of $\omega$ with $t=1$ involves a
gauge parameter $\lambda$ with $t=2$ associated to the extra field
$\omega$ with $t=2$, cf. \eqref{gaugeparameters}. We now propose
the extended action functional that contains fields associated to
all coordinates of the target superspace  $\manM$ \eqref{Ms} on
equal footing.

It is convenient to represent elements of  $\manM$ \eqref{Ms} as polynomials $\omega(y,p)$ in auxiliary
variables  $y^a$ and  $p_a$. The irreducibility conditions imposed on expansion coefficients of
$\omega(y,p)$ are encoded by
\begin{equation}
\begin{gathered}
\label{A-const}
 p^a\dl{p^a}\,\omega=(s-1)\omega\,, \qquad  p^a\dl{y^a}\, \omega=0\,,\\ ~~~\dl{p^a}\dl{p_a}\, \omega=0\,, \qquad \dl{y^a}\dl{p_a}\, \omega=0\,, \qquad \dl{y^a}\dl{y_a}\,\omega=0\,.
\end{gathered}
\end{equation}
Expanding $\omega$ into homogeneous in $y$ components one finds precisely $s$
coordinates $\omega_0 \ldots, \omega_{s-1}$ identified with higher spin connections \eqref{Ms}.

In what follows we denote by $\inner{}{}$ the natural inner product on the space of polynomials in $y,p$,
\textit{i.e.} one determined by $\inner{y^a}{y^b}=\eta^{ab}$ and $\inner{p^a}{p^b}=\eta^{ab}$.  To describe the flat
background one also introduces extra coordinates (they are external parameters from the point of view of $\manM$)
encoded in  $e_0=e^a p_a$ and $\omega_0=\omega_0^a{}{}_b y^b p_a$. For simplicity
put $\omega_0=0$ but the nontrivial $\omega_0$ can always be reinstated.

The space $\manM$ can be  extended by extra coordinate $R_{ab}(y,p)$ of ghost degree $0$ and
homogeneity $s-1$ in the auxiliary variables  such that it is  a rank $2s$ Lorentz tensor with the symmetry of $o(d-1,1)$ rectangular two-row Young diagram. In addition, it is required to be traceless in the sector of $y,p$ variables.
The conditions can be summarized as
\begin{equation}
\label{R}
\big(p_c\dl{p_c}-s+1\big)R_{ab}=0\,,
\quad  \dl{p_c}\dl{p^c} R_{ab}=0\,,\quad p_c\dl{y_c} R_{ab}=0\,, \quad\dl{p^{[a}}R_{bc]}=0\,,
\end{equation}
where we only list the minimal set from which the remaining ones follow as consistency  conditions. Note that for $s>2$
it follows that coefficients of $R_{ab}(y,p)$ are totally traceless in $2s$ indices  while
their tensor structure is precisely that of the spin $s$ Weyl tensor. In the case of $s=2$ the second
condition in \eqref{R} is satisfied trivially so that the respective tensor has properties of the Riemann curvature.

The resulting supermanifold is equipped with the following odd nilpotent  vector field $Q$ determined by
\begin{equation}
\label{Qomega}
 Q\omega=-e_0^a\dl{y^a}\omega+e_0^a e_0^b R_{ab}\,, \qquad Q R_{ab}=0\;.
\end{equation}
In terms of components
\begin{equation}
 Q\omega_i=-e_0^a\dl{y^a}\omega_{i+1}\qquad i=0,...,s-2\,,\qquad Q\omega_{s-1}=e_0^a e_0^b R_{ab}\;.
\end{equation}
We  go further from  $\manM$ to $\hat\manM$ which is a cotangent bundle $T^*[n-1]\manM$ extended by $R_{ab}$. The coordinates on the fibres carry
degree $n-2$ and are denoted by $\Lambda$. It is convenient to encode the fiber coordinates into $\Lambda(y,p)$
satisfying~\eqref{A-const}.

Let $\hat\manM$ be equipped  with the following $1$-form and the potential (the potential is taken from~\cite{Grigoriev:2012xg})\comm{Note that at this stage one can in principle take as $\hat L$ any $Q$-invariant functions of $A$.}
\begin{equation}
\begin{gathered}
\chi=\inner{d \omega}{\Lambda }\,, \qquad \SM=\chi(Q)-\HM\,, \\
\HM= -\half \bar\Vol^{ab}\big(\inner{\omega_{1a}}{\omega_{1b}}-\inner{p_a \omega_{1c}}{p_b \omega_1^c}\big)\;,
\end{gathered}
\end{equation}
where $\omega_{1a}$ is defined through $\omega_1(y,p)=y^a \omega_{1a}(p)$, and the potential
$\HM=  \HM (e_0,\omega_1)$ is $Q$-invariant,
\begin{equation}
\label{QL}
Q \HM= 0 \;.
\end{equation}

The odd vector field $Q$ is extended to $\Lambda$-variables as follows
\begin{equation}
\label{Qlambda}
 Q\Lambda=\cP [e_0^a\, y_a\, \Lambda] + q\;,
\end{equation}
where $\cP$ denotes the projector to the subspace~~\eqref{A-const} and $q$ is uniquely determined by
\begin{equation}
\label{q}
\inner{d \omega}{q}=-d\omega_i \dl{\omega_i} \HM\,.
\end{equation}

All the variables save for $\Lambda_{s-1}$ and $R_{ab}$ are vertical and hence have
their associated gauge parameters.
The introduced $Q,\chi,\SM$ satisfy the  compatibility conditions \eqref{c1} - \eqref{c3}.
To see this let us restrict to $s>2$ first (the spin 2 case is treated explicitly below).
In the first step we set $\HM=0$ and find that the compatibility conditions are fulfilled provided the  extra coordinate
$R_{ab}(y,p)$ satisfies \eqref{R}.


Turning on a non-vanishing $\HM (e_0,\omega_1)$ the extra contributions in the first compatibility
condition  is
\begin{equation}
 \inner{q}{d\omega}=-d\HM
\end{equation}
which is a definition of $q$ \eqref{q}. And
\begin{equation}
\inner{q}{Q\omega}=i_Q \inner{q}{\dps d\omega}= - i_Q d \HM =- Q\HM=0
\end{equation}
again thanks to the definition of $q$. In the last equality we used the $Q$-invariance property
of $\HM$ \eqref{QL}.

It follows that the frame-like action \eqref{masteraction} determined by $Q,\chi,L$ is consistent. It is given explicitly
by
\begin{multline}
\label{extaction}
\dps
S_{0}[\omega, \Lambda, R] = \int
\sum_{i=0}^{s-1}  \inner{\derham \omega_i-  e_0^a\dl{y^a}\omega_{i+1}
+\delta_{i,s-1}e_0^a e_0^b R_{ab} }{\Lambda_i }
\\
\dps
+\half {\bar\Vol}^{ab}\big(\inner{\omega_{1a}}{\omega_{1b}}-\inner{p_a\, \omega_{1c}}{p_b\, \omega_1^c}\big)\;.
\end{multline}
Combinations of terms  $\derham \omega_i- e_0^a\dl{y^a}\omega_{i+1} \equiv R_i$ are in fact linearized curvatures
associated to fields $\omega_i$.
Note that for  $i=s-1$ curvature $R_i$ enters action \eqref{extaction} together with the  extra
coordinate $R_{ab}$. For $i \neq s-1$ one also represents $R_i$ as $R_i=(\derham+Q)\omega_i$ so that
the first terms in \eqref{extaction} can be written as $\inner{(\derham+Q)\omega_i}{\Lambda_i}$.
At the same time, the potential term $\HM$ depends on fields $\omega_1$ only.

Before showing that the above action indeed describes spin-$s$
massless field let us discuss the interpretation of the various structures
it involves. First of all, the odd vector field $Q$ considered on
$\manM$ with coordinates $\omega_i$ is directly related to the
linear operator $\sigma_-=e_0^a \dl{y_a}$ defined on the space of
polynomials in auxiliary variables $y,p,e_0$ seen as differential
forms with values in \eqref{A-const}.   The operator
$\sigma_-$ and its cohomology are well-known within  the unfolded
description of higher spin  fields~\cite{Lopatin:1988hz,Bekaert:2005vh}. In the present
consideration we prefer to work with $Q$ rather than $\sigma_-$ because
it is naturally defined on generic function(als) of fields. In
particular, potential $\HM$ is naturally a $Q$-cocycle in the
sense that it is $Q$-closed \eqref{QL} and adding a $Q$-exact term leads to
an equivalent action. Finally, the operator $\cP(e^ay_a)$ entering
\eqref{Qlambda} is conjugated to $\sigma_-$ and is also known in
the unfolded approach as $\sigma_+$.

%
%
%
%

The extended frame-like action \eqref{extaction} is equivalent to
the conventional frame-like action \eqref{linHS}. To show this one
observes that $\Lambda_{s-1}$ and $R_{ab}$ are auxiliary fields
that can be expressed in terms of other fields using their own
equations of motion. Indeed, using the gauge symmetry for
$\Lambda_{s-1}$ one can eliminate all the components of the image
of $\cP(e^ay_a)$, while the remaining components are precisely one
to one with the components of $R_{ab}$. The same applies to those
components of $\omega_i$, $i=2,\ldots, s-1$ and $\Lambda_{i}$,
$i=1,\ldots, s-2$ that  cannot be eliminated using the gauge
transformations for these fields. In particular, varying the
Lagrange multipliers $\Lambda_i$, $i=1,\ldots, s-2$  one arrives at
the zero-curvature constraints $R_i =0$, $i =
1,\ldots ,s-2$.\,\footnote{It is worth mentioning that implementation
of zero-curvature constraints through the frame-like actions with
Lagrange multipliers have been discussed in
\cite{Vasiliev:1988sa,Vasiliev:2009ck}}. Finally, upon elimination one ends up
with the reduced action depending on $\omega_{0}, \omega_1$ and $\Lambda_0$
and given by the \eqref{extaction} where all the other fields are
put to zero. But this is precisely the action which was shown
in~\cite{Grigoriev:2012xg} to produce frame-like
action~\eqref{linHS}.

A crucial point in the above argument is that all the components
of $R_{ab}$, $\omega_i$, $i=2,\ldots, s-1$ and $\Lambda_{i}$,
$i=1,\ldots, s-1$ can be eliminated using the gauge symmetries and
the equations of motion. This fact can be traced to the properties of
the cohomology of $\sigma_-=e^a\dl{y^a}$ entering the first term
in~\eqref{extaction}. More precisely,  1-forms $\omega$ can be
identified with the linear in $e_0^a$ elements of polynomials in
$e_0^a,p^a,y^a$ satisfying~\eqref{A-const} and similarly for
fields $\Lambda_i$. One can actually show that all the components
of $\omega_i$ except $\sigma_-$ cohomology and all the components
of $\Lambda_i$ except $(\sigma_-)^\dagger=\sigma_+$-cohomology are generalized
auxiliary (see, \textit{e.g.}, \cite{Grigoriev:2012xg} for further
details on generalized auxiliary fields) if one disregards $\HM$ and $R_{ab}$.  Restricting to
$i>1$ the only relevant $\sigma_-$ cohomology class  is the Weyl
tensor in degree 2 and homogeneity $s-1$ in $y$ along with its
conjugate $\sigma_+$ class in degree $n-2$. Because $\omega_i$ are
of degree $1$ all $\omega_i$ with $i>1$ are not in the cohomology and can be
eliminated. In contrast, the component of $\Lambda_{s-1}$
corresponding to $\sigma_+$ cohomology can not because $\Lambda$
is of degree $n-2$. However, in the expression for the action this
component enters multiplied by $R_{ab}(y,p)$ so that both the component and $R_{ab}$
are auxiliary fields. In other words from this perspective
$R_{ab}$ plays a role of Lagrange multiplier needed to put to zero
the unwanted component of $\Lambda_{s-1}$. Alternatively, $R_{ab}$
can be seen as an element eliminating (gluing) $Q$-cohomology in the space of
linear in $\omega_{i}$ functions on $\manM$. These arguments can
be seen as Lagrangian counterpart of the $\sigma_-$-cohomology
method~\cite{Bekaert:2005vh} at the level of equations of motion.

Furthermore,
one can generally allow new $R_{ab}$-type fields to enter the
terms proportional to $\Lambda_i$ with $i<s-1$. A particularly
interesting option is to introduce $F_{ab}(y,p)$ in the term with
$\Lambda_1$ (for consistency, one also needs to introduce its
descendants in the terms proportional to $\Lambda_i$ with $i>1$). This is the well-known trick to
relax the Fronsdal equations because such $F_{ab}$ precisely
corresponds to Fronsdal tensor. In its turn $F_{ab}$ is again
related to $\sigma_-$ cohomology class known in the unfolded
approach as ``Einstein'' cohomology.

Let us finally mention that the action~\eqref{extaction} can be
generalized to describe spin-$s$ massless fields on the (A)dS
background. In this case one needs to allow  nonvanishing
background $\omega_0$ and add $\Lambda\sigma_+ = \Lambda\cP (e_0^a\, y_a)$
contribution to the first term, where $\Lambda$ is the cosmological constant.
The odd vector field  $Q$ acting on $\hat\manM$ will be augmented by $\Lambda\sigma_+$ and $\Lambda\sigma_-$
for coordinates $\omega(y,p)$ and $\Lambda(y,p)$, respectively.

\vspace{-3mm}

\paragraph{The spin $s=2$ case.} Let us explicitly list all the structures in the case of spin $2$.
The coordinates on $\hat\manM$ are introduced according to
\begin{equation}
\omega_0=e^a p_a\,, \quad \omega_1=\omega^a{}_b y^b p_a\,, \quad \Lambda_0=\lambda_a p^a\,,
\quad \Lambda_1=\lambda_{ab}y^a p^b\,, \quad R_{ab}=R_{ab}^{cd}\, y_c p_d\;.
\end{equation}

Components of the odd vector field  $Q$ are given by
\begin{equation}
\begin{gathered}
 Qe^a=\omega^{a}{}_b e_0^b\,, \qquad  Q \omega^{ab}=e_0^c e_0^d R_{cd}^{ab} \,,\qquad QR_{ab}^{cd}=0\,,\\
 Q \lambda_a=0\,,\quad Q \lambda_{ab}=\lambda_{[a} e_0{}_{b]}+\omega_{[a}{}^c{\bar\Vol}_{c]b}\;.
\end{gathered}
\end{equation}
The 1-form $\chi$ is
\begin{equation}
 \chi=\lambda_a de^a+\half\lambda_{ab} d\omega^{ab}\;,
\end{equation}
and the potential $\SM$
\begin{equation}
 \SM=\chi(Q)+\half \bar\Vol_{ab}\,\omega^{a}{}_{c}\omega^{cb}\,.
\end{equation}
All the variables are vertical except  for $\lambda_{ab}$ and $R$. In particular, there are gauge parameters
$\xi^a,\xi^{ab},\epsilon_a$ associated to respectively $e^a,\omega^{ab},\lambda_a$.
One can explicitly check that the compatibility conditions \eqref{c1}-\eqref{c3}  are fulfilled. Namely,
\begin{equation}
 i_Q \sigma-d\SM=\lambda_{ab}e_0^c e_0^d dR^{ab}_{cd}\,, \qquad \sigma(Q,Q)=0\,, \qquad Q\SM=0\,.
\end{equation}
In checking these relations we made use of the following consequences of Young symmetries of $R^{ab}_{cd}$:
\begin{equation}
 R^{ab}_{cd}e_0^c e_0^d e_0{}_a\equiv 0\,, \qquad \omega^a{}_b R^{bc}_{ac}\equiv 0\,.
 \end{equation}

The linearized gravitational action takes the form
\begin{equation}
\label{actionDRAFT}
S_0[e,\omega, \lambda, R]=\int \lambda_a(\derham e^a +\omega^{a}{}_b e_0^b)+\lambda_{ab}(\derham \omega^{ab}+e_0^c e_0^d R^{ab}_{cd})
 + \frac{1}{2}\, \bar\Vol_{ab}\,\omega^{a}{}_{c}\omega^{cb}\;.
\end{equation}
The theory \eqref{actionDRAFT} can be made  non-linear just by replacing the background fields $e_0$ and
$\omega_0 =0$ with dynamical fields $e$ and $\omega$. Note that one also needs to reinstate the familiar
$\omega^a{}_c\omega^{cb}$ term in the expression for the curvatures in the second term. The resulting action will describe the standard Einstein gravitation theory action equivalent to \eqref{actgravity}.

\section{(Polymomentum) Hamiltonian description}
\label{sec:poly}
\subsection{ One-dimensional  constrained Hamiltonian systems}
\label{sec:onedim}

It is well-known that equations of motion of generic one-dimensional  Lagrangian system can be rewritten in a
presymplectic Hamiltonian form, \textit{e.g.} through hamiltonization. Moreover, the extended
Hamiltonian action of a constrained system can be represented in the AKSZ form where
the extended phase space of BFV-BRST formulation
plays the role of the target space ~\cite{Grigoriev:1999qz}. Whence, the system is frame-like one. The presymplectic
representation of the Euler--Lagrange equations can be seen even without resorting to constrained
Hamiltonian formalism. The easiest way to observe this for the Lagrangian $L(q,\dot q)$ is to introduce auxiliary fields
$v^i=\dot q^i$ and $p_i=\ddl{L(q,v)}{v^i}$ so that the system is equivalently represented as \cite{Gitman:1990qh}
\begin{equation}
S[p,q,v]= \int dt (p(\dot q -v)+L(q,v))\,.
\end{equation}
Indeed, eliminating variables $p$ and $v$ by  their own equations of motion gives back the starting point
Lagrangian. At the same time, the equations of motion determined by $S$ have the presymplectic Hamiltonian form
\begin{equation}
 \dot \Psi^A \sigma_{AB}+\d_B H=0\,,
\end{equation}
where $\Psi^A=(q^i,p_i,v^i)$, $H:=p_i v^i -L(q,v)$ and $\sigma:=dp_i \wedge dq^i$.  Using the system as a
starting point of the presymplectic version~\cite{Gotay:1978,Gotay:1978dv} of the Dirac--Bergmann algorithm one
ends up with the reduced presymplectic system.  This can equivalently be obtained using conventional
Dirac--Bergmann approach: the reduced presymplectic system is just the constrained surface equipped with the
pullbacks of the phase space symplectic form and the Hamiltonian.

We now take a different route and analyze what do general axioms of Section~\bref{sec:relax} tell us in the
one-dimensional  case. It turns out that constrained systems appear from a purely supergeometrical perspective without introducing the full-scale BFV-BRST formulation.
Restricting to systems in $n=1$ dimension we take $\manM$ to be a presymplectic manifold with bosonic coordinates $\psi^i$ and fermionic $C^A$  of ghost degree $0$ and $1$ respectively. We assume ghost variables $C^A$ split into vertical $c^\alpha$ and horizontal $c^a$ while $\phi^i$ are also assumed horizontal.  Taking into account the ghost degree, $\chi$ has the form $\chi=d\psi^i \chi_i(\psi)$. Similarly,  $Q$ of degree $1$ reads as $Q\psi^i=R^i_\alpha c^\alpha+R^i_a c^a$ and $Qc^A =-\half U^A_{BC} C^B C^C$ where $C^A=\{c^\alpha,c^a\}$. For simplicity we assume $R^i_\alpha$ to be of maximal rank and $Q$ nilpotent. In particular, $\commut{R_A}{R_B}=U_{AB}^C R_C$.
A generic expression for $\SM$ is $\SM=c^\alpha T_{\alpha}(\psi)+c^a T_a(\psi)$. The compatibility conditions~\eqref{c1}-\eqref{c3} say
\begin{gather}
 \qquad \d_i T_\alpha-\sigma_{ij}R^j_{\alpha}=0\,,\qquad L_{R_\alpha}\sigma=0\,,\\
 \label{RT-cons}
  \half(R_\alpha T_{\beta}-R_\beta T_\alpha)=U_{\alpha \beta}^C T_C\,,          \qquad
  R_\alpha T_b=U_{\alpha b}^C T_C\,.
\end{gather}
A frame-like action determined by this data reads as
\begin{equation}
S[\psi, c]=\int (\derham \psi^i \chi_i +\SM)=\int d\tau(\chi_i \dot \psi^i+c^\alpha T_{\alpha}+c^a T_a)\;,
\end{equation}
where $c^\alpha,c^a$ are now $1$-forms and are to be identified with Lagrange multipliers.
The gauge transformations take the form:
\begin{equation}
 \delta_\lambda \psi^i=-R^i_\beta \lambda^\beta\,, \qquad \delta c^\alpha =\derham \lambda^\alpha +\lambda^\beta U^{\alpha}_{\beta B} C^B\,, \qquad  \delta c^a =\lambda^\beta U^{a}_{\beta B} c^B\,.
\end{equation}
It is clear from the structure of the action and the gauge transformations that we are dealing with a generalization
of a constrained Hamiltonian system where $T_\alpha$ and $T_a$ play the role analogous to the first and the
second class constraints, respectively.

Under suitable regularity assumptions one can solve equations of motion $T_a=0$ so that $\manM$ is replaced
with the submanifold $\manM^\prime \subset \manM$ singled out by $T_a=0$ and $c^a=0$ and we take
$\psi^\mu$ to be the independent coordinates on $\manM^\prime$. More precisely, taking  $T_a$
as a part of the coordinate system on $\manM$ one finds that $T_a$ and $c^a$ give rise to auxiliary fields. Their elimination
results in the following action
\begin{equation}
  S[\psi,c]=\int \derham \psi^\mu \chi_\mu +c^\alpha T_{\alpha}\,.
\end{equation}

The subtlety is that $Q$ is in general not tangent to
$\manM^\prime$ (this happens if $Qc^a$ and $QT_a$ do not vanish when $T_a=C^b=0$)
so that its gauge symmetry can not be easily represented in terms of $Q$. Assuming for simplicity
$Q$ and $\SM$ are such that $Q$ is tangent to $\manM^\prime$ the expression for the gauge transformations
take the usual form
\begin{equation}
 \delta_\lambda \psi^\mu=-R^i_\beta \lambda^\beta\,, \qquad
 \delta c^\alpha =\derham \lambda^\alpha +\lambda^\beta U^{\alpha}_{\beta \gamma} c^\gamma\,.
\end{equation}
This system is nearly a usual first-class constrained Hamiltonian system. The only difference is that $\sigma$
is not necessarily invertible. If one assumes $\sigma_{\mu\nu}$ invertible one finds  $R^\mu_\alpha=\sigma^{\mu\nu}\d_\mu T_\alpha$ so that $R_\alpha=\pb{T_\alpha}{\cdot}$, where $\pb{\cdot}{\cdot}$ is a Poisson bracket determined by $\sigma$ on the space of $\psi^\mu$-variables so that indeed we are dealing with the conventional first-class constraint system.

If one stays in the symplectic framework an extra equivalent reduction is also possible. Namely,
under the suitable regularity assumptions one can take $T_\alpha$ as a part of the coordinate system. Just like in the previous step variables  $c^\alpha$ and $T_\alpha$ are auxiliary and can be eliminated. The action of the reduced system takes the form
(here $\psi^m$ are coordinates on $\manM^{\prime\prime}$ singled out by $T_\alpha=0$)
\begin{equation}
\label{final-red}
 S[\psi]=\int \derham \psi^m \chi_{m}\;,
\end{equation}
while its gauge symmetries are determined by $Q$ restricted to the surface $\manM^{\prime\prime}\subset \manM^\prime$.

It is important to note that in contrast to the previous
reductions this one, strictly speaking, takes us outside the class
of systems described in Section~\bref{sec:relax}.  Indeed,
although the 1-form field (Lagrange multiplier) associated to
coordinate $c^\alpha$ has been eliminated using equations of
motion the gauge invariance with parameter $\lambda^\alpha$ is
still present. This is exactly the same situation as we met in the
case of the frame-like Lagrangians for HS fields considered in Section
\bref{sec:HS}: there are ghost coordinates on $\manM$ that possess
associated gauge parameters but do not possess associated
fields (or, equivalently, the respective fields do not enter a given
frame-like Lagrangian).

The following remarks are in order:

-- the system has a vanishing Hamiltonian. To include the Hamiltonian one needs to treat one of the ghosts
$c^\alpha$ (say, $c^0$) as a horizontal variable and the respective field as a background einbein field. A different way to introduce a genuine Hamiltonian is to reinterpret the system as a parameterized Hamiltonian system (for more details see the discussion in Section~\bref{sec:param}).

-- The example is restricted to the case of irreducible systems. To describe reducible ones one allows for ghost variables of degree $2$ (and higher) and lets $Q$ to encode reducibility relations.

-- We have concentrated on geometry and gauge symmetries and have not discussed the dynamical
implications~(see, \textit{e.g.},~\cite{Alkalaev:1999hi,Saavedra:2000wk} and references therein) of degenerate symplectic structures.
\comm{looks like only irregular case is interesting as regular merely reduces to usual constraints.}



\subsection{Parameterized systems}
\label{sec:param}
By construction frame-like Lagrangians describe diffeomorphism-invariant systems. In the above examples the non-diffeomorphism invariant theories
are described by coupling them to the vacuum gravitational field. There is a different approach to describing such theories based on parametrization.
It is well known~( see e.g.~\cite{Arnowitt:1959ah,Kuchar:1976yy,Torre:1992bg}) that any theory can be made parametrization invariant by introducing extra fields
and extra gauge transformations. More precisely,
if $L[\phi,\d_a\phi]$ is a Lagrangian of a system with fields $\phi^i$ let us consider a new system with fields $\phi^i,z^a$ depending on generic space time coordinates $x^\mu$.
The parameterized action is given by
\begin{equation}
 S[\phi,z]=\int d^n x \det(e)L[\phi,e^\mu_a \d_\mu \phi]\,, \qquad e^a_\mu\equiv \ddl{z^a}{x^\mu}\,,\quad e^a_\mu e^\nu_b=\delta^a_b\,.
\end{equation}
It is invariant under the following gauge transformation:
\begin{equation}
 \delta_\lambda z^a=\lambda^a\,, \qquad \delta \phi=\lambda^a e_a^\mu \d_\mu \phi\,,
\end{equation}
where $\lambda^a$ are components of a vector field parameterizing infinitesimal diffeomorphisms.

It turns out that parameterized systems can naturally be written in the frame-like form. Let us begin with the scalar field example. In addition to coordinates $\phi,\pi^a$ introduced in Section~\bref{sec:gravity+scalar} we need the following coordinates on $\manM$: $z^a,e^a,p_a,\varrho_a$ whose degrees are respectively $0,1,n-1,n-2$. The differential $Q$ is given by
\begin{gather}
\label{Qparam}
 Qz^a=-e^a\,, \quad Q e^a=0\,, \quad  Q p_a=0\,, \quad Q \varrho_a=-   p_a\,, \\
 Q\phi=Q\pi^a=0\,.
\end{gather}
The 1-form $\chi$ is given by
\begin{equation}
 \chi=dz^a p_a +de^a \varrho_a  + d\phi \pi^a  \Vol_a\;.
\end{equation}
Note that the last term coincides with the respective contribution in the 1-form of the scalar in a gravity background.
Taking $\HM$ as in Section~\bref{sec:gravity+scalar} results in the following action
\begin{equation}
S[z,p,e,\varrho,\pi,\phi] = \int \big[ (\derham z^a-e^a)p_a +\derham e^a \varrho_a  +\derham \phi \pi^a \Vol_a+\half  (\pi^b\pi_b- m^2\phi^2)\big)\Vol]\,.
 \end{equation}
Let us explicitly  spell out its gauge symmetries (we only present nonvanishing transformations)
\begin{equation}
 \delta z^a=\xi^a\,, \qquad \delta \varrho_a=\epsilon_a\,,
\end{equation}
where $\xi^a$ is the 0-form parameter associated with vertical coordinate $e^a$ while $\epsilon_a$ is the $n-2$-form parameter associated to vertical coordinate $p_a$. Taking into account the equations of motion following from the above action (we only spell out those in the sector of $z,e,p,\varrho$\,-variables)
\begin{equation}
 e^a-\derham z^a=0\,, \quad \derham e^a=0\,, \quad p_a=U_a(e,\phi,d\phi,\pi,\varrho)\,,
\end{equation}
it is clear that variables $z^a,\varrho_a$ are St\"ueckelberg while $e^a,p_a$ are auxiliary so that all of them do not bring in new degrees of freedom. Indeed, these gauge symmetries are enough to put e.g. $z^a=x^a$ and $\varrho_a=0$. The equations of motion then say $e^a=dz^a=dx^a$ and fix $p_a$ in terms of the remaining variables. Upon gauge fixation and elimination of
auxiliary fields the action becomes just~\eqref{scalar-1st}.

Let us stress the difference between the above parameterized formulation and the scalar field action in the flat gravity background described by $e_0^a,\omega_0^{ab}$. In the gravity case fields $e_0,\omega_0$ are to be treated as background fields. In contrast, in the above parameterized system all the fields enter the Lagrangian at the equal footing. These are equations of motion and gauge symmetries which make the additional fields non-dynamical.

More generally, if one is given with the frame-like action involving flat gravity background described by $e_0^a, \omega_0^{ab}$  one can systematically rewrite it in  the parameterized form. Indeed, suppose that the frame-like description is provided by the action of the form
\begin{equation}
 S[\Psi]=\int \derham \Psi^A \chi^0_{A}(\Psi,e_0,\omega_0)+\SM(\Psi,e_0,\omega_0)\;,
\end{equation}
where $e_0$ and $\omega_0$ is the flat gravity background such that $\derham e_0+\omega_0 e=0$, $\derham \omega_0+\omega_0\omega_0=0$ and gauge symmetries are determined by $Q_0$. Introducing variables $z,e,p,\varrho$ in exactly the same way as
in the scalar field example one takes
\begin{equation}
 \chi=\chi^0+dz^a p_a +de^a \varrho_a\;,
\end{equation}
where in $\chi_0$ in the RHS one puts $\omega_0$ to zero and replaces $e_0$ with $e$.  Taking $Q$ as in \eqref{Qparam} in the sector of new variables and unchanged in the original sector one ends up with the parameterized description:
\begin{equation}
S[z,p,e,\varrho,\Psi]=\int (\derham z^a-e^a) p_a +\derham e^a \varrho_a +\derham \Psi^A \chi^0_A(\Psi,e)+\SM(\Psi,e)\,.
\end{equation}
In this Lagrangian all fields can be treated as non-background. Of course in the gauge $z^a=x^a$ it is equivalent to the starting point system.

The following toy example illustrates that the above parametrization procedure reduces to the usual parameterized Hamiltonian description in the case of a Hamiltonian system. To see this let us take as $\manM$ a phase space of a Hamiltonian system with Hamiltonian $H(\psi)$ and symplectic potential $\chi=d\psi^a \chi_a(\psi)$ and extend it by odd degree $1$ coordinate $e$ (einbein) so that the Hamiltonian action is $\int \derham \psi^a \chi_a-eH$. Following the above procedure and introducing variables $z,p,\varrho$ the parameterized action reads as
\begin{equation}
 S[\psi,e,z,p]=\int (\derham \psi^a \chi_a + \derham z p-e(H+p))\,,
\end{equation}
(note that the term $\derham e \varrho$ is missing as $\gh{\varrho}=-1$ so that there are no
dynamical fields associated to $\rho$) which is a standard Hamiltonian action of a usual parameterized
Hamiltonian system (see, \textit{e.g.},~\cite{HT-book}).

It is instructive to write down an analog of \eqref{final-red} in this case. The constrained surface is parameterized by $\psi^a,t:=z$ and the pullback of the 1-form is $d\psi^a\chi_a+H(\psi)dt$ so that the reduced action takes the well-known invariant form
\begin{equation}
 S[\psi,t]=\int d \tau  (\derham \psi^a\chi_a+H dt)=\int \Phi^*(d\psi^a \chi_a+H dt)\,,
\end{equation}
where $\Phi^*$ denotes the pullback induced by the map $\psi^a(\tau),t(\tau)$
from a time line to the constrained surface with coordinates $\psi^a,t$.

To complete the description of parameterized system let us mention that the structure of the Lagrangian and gauge transformations in the sector of $z,e,p,\varrho$ variables explicitly coincides with that of the parent Lagrangians from~\cite{Grigoriev:2010ic,Grigoriev:2012xg}. This is not a coincidence as frame-like formulation of a given (say in the metric-like formalism) system can be systematically derived from the parent formulation. Mention also that the above parameterized description does not contain $\omega_\mu^{ab}$ field. In fact this only applies to the minimal version. If for instance the starting point system is Lorentz invariant (this is of course always true if the system originates from that on a flat gravity background) one can systematically gauge this symmetry which results in $\omega_\mu^{ab}$ field present in the formulation. Analogous extension was discussed in the context of parent formulation in~\cite{Grigoriev:2012xg,Bekaert:2012vt}.

\subsection{Relation to the polymomentum phase space}
\label{sec:polurel}

Given a frame-like system with the target supermanifold $\manM$ equipped with $Q,\chi,\SM$ and the ghost degree
let  $\manM_P$ be the space  of independent variables $x^\mu$ and dependent ones $\Psi^A_{\mu_1\ldots\mu_{p_A}}(x)$.
Since the formulation under consideration is the first-order,   $\manM_P$ can be visualized as the multidimensional analog
of the phase space. Moreover, as we are going to see $\manM_P$ can be merely identified
with the polymomentum phase space of the de Donder--Weyl formalism. More precisely, we now show how the basic structures of the polymomentum approach arise from the frame-like formalism developed above. A systematic exposition of the polymomentum approach can be found in, \textit{e.g.},~\cite{Gotay:1997eg,Kanatchikov:2000jz}.


%

Space $\manM$ is equipped with $1$-form $\chi$ of ghost degree $\gh\chi = n-1$. This form gives rise to an $n$-form
on the phase space $\manM_P$. Indeed, substituting $\Psi^A$ in the expression for $\chi$ with $\Psi^A_{\mu_1\ldots\mu_{p_A}}\,dx^{\mu_1}\ldots dx^{\mu_{p_A}}$ one arrives at $n$-form $\chi_P$ defined on $\manM_P$. This form can be identified with a version of the canonical $n$-form of the polymomentum approach, see, \textit{e.g.},~\cite{Gotay:1997eg,Kanatchikov:2000jz}.

$n$-form $\chi_P$ is in turn related to the canonical 1-form of the usual Hamiltonian formulation. Indeed, explicitly separating the space-time into the space $\Sigma$ with coordinates $x^i$, $i=1,\ldots, n-1$ and time $x^0$ and integrating $\chi_P$
over the space-like $(n-1)$-dimensional hypersurface $\Sigma$ determined by $x^0=\const$ gives the following
canonical 1-form
\begin{equation}
\label{can-1f}
 \int_\Sigma  \delta\Psi^A  \wedge \chi_A(\Psi)\,.
\end{equation}
It is the functional $1$-form on the space of fields $\Psi^A_{i_1\ldots i_{p_A}}$
defined on $\Sigma$.

To see that \eqref{can-1f} is the usual 1-form of the Hamiltonian
formalism observe that frame-like action \eqref{masteraction} takes the form
\begin{equation}
S[\Psi]=\int dx^0 \int_\Sigma \frac{d}{dx^0}\Psi^A \chi_A(\Psi)+ \int dx^0 \int_{\Sigma}\derham_S\Psi^A\chi_A(\Psi)+\SM(\Psi) \;,
\end{equation}
where $\derham_S= dx^i\dl{x^i}$ is the spatial part of the de Rham differential.

As an illustration let us consider the scalar field and the Yang-Mills theory examples.
For simplicity we take Minkowski spacetime as the background described in Cartesian coordinates $x^a$ by
$e^a=\derham x^a$ and $\omega^{ab}=0$.

In the case of the scalar field theory the respective component
of the $n$-form $\chi_P$ is given by
$\chi_P=\pi^a \Vol_a$. Integrating over space $\Sigma$ gives the usual canonical $1$-form
$\dps\int_\Sigma\, \pi^0 \, \delta\phi$, and, as expected, $\pi^0$ is the usual momenta.

In the case of the Yang-Mills theory  one obtains
$
\dps\chi_P=\left< F^{ab}, \dm A_{a}\right> \Vol_b
$.
This can be identified with that of the canonical $n$-form of the covariant
Hamiltonian formalism (see, \textit{e.g.},~\cite{Gotay:1997eg}). After integrating over space $\Sigma$ one arrives at
$
\dps \int_\Sigma \left<F^{a0}, \delta A_a\right>
$,
so that again, as expected, $F^{i0}$ with $i=1,..., n-1$ is the usual canonical momenta conjugated to the
spatial components $A_i$ of the Yang-Mills  field.

It follows from the above considerations that the structure of the polymomentum phase space as well
as the usual phase space is completely determined by that of the supermanifold $\manM$. This suggests that
$\manM$ is more fundamental object that can be used as a substitute of the polymomentum phase space. Note that
from this perspective frame-like actions considered in this work are multidimensional generalizations
of the extended Hamiltonian actions for constrained systems in 1d.

Since we deal with a gauge theory the definition of the field  space
or phase space is ambiguous. More precisely, one can always add extra fields along with extra terms in the Lagrangian or
the extra gauge transformations so that these extra variables do not bring in new degrees of freedom. Such variables
are known as auxiliary and/or St\"uekelberg fields. In particular, according to Section~\bref{sec:param} one can always
consider a parameterized version of the theory, where the spacetime coordinates are part of the target space $\manM$
(as it is usually the case for the standard~\cite{Gotay:1997eg,Kanatchikov:2000jz} polymomentum description of theories not coupled to gravity).

\appendix

\section{Notation and conventions}
\label{sec:appendix}

Let $\manM$ be a supermanifold with local coordinates $z^A$. Vector fields are left derivations
of the algebra of smooth functions $\cC_{\manM}$ on $\manM$,  \textit{i.e.}, $V:C_\manM \to C_\manM$ is a linear map satisfying
$V(fg)=(Vf)g+(-1)^{\p{V}\p{f}}f(Vg)$, where $|\cdot| = 0,1$ denotes a parity. Vector fields clearly form a left module over $\cC_\manM$.

In order to work with differential forms it is useful to consider the odd tangent bundle $\Pi T \manM$
over $\manM$. Differential forms are simply functions on $\Pi T \manM$ which are polynomials in the fiber coordinates.
By slight abuse of notation, the fiber coordinates are denoted by $dz^A$. The components of a $p$-form
$\alpha$ are introduced as follows:
\begin{equation}
\alpha(z,dz)=\frac{1}{p!}\,dz^{A_p}\ldots dz^{A_1}\alpha_{A_1\ldots A_p}(z)
\end{equation}
It follows that in our conventions the graded antisymmetry property of components
is determined by $dz^Adz^B=(-1)^{(\p{A}+1)(\p{A}+1)}dz^Bdz^A$. In particular, for the components of a
$2$-form $\sigma$ one has
\begin{equation}
\sigma_{AB}=(-1)^{(\p{A}+1)(\p{B+1})}\sigma_{BA}\,.
\end{equation}

De Rham differential is a natural vector field on $\Pi T\manM$ given by
\begin{equation}
 \dm=dz^A\dl{z^A}\,,
\end{equation}
where $\dl{Z^A} = \d_A$ denotes a left derivative.
For instance, for $0$-forms $z^A$ and $1$-form $\chi=dz^A \chi_A(z)$ one explicitly gets
\begin{equation}
\dm z^A=dz^A\,, \qquad \dm (dz^B \chi_B)=\frac{(-1)^{\p{B}+1}}{2}dz^Bdz^A(\d_A\chi_B-(-1)^{\p{A}\p{B}}\d_B\chi_A)\,.
\end{equation}
In terms of components,
\begin{equation}
\label{2formcomp}
(\dm\chi)_{AB}=(-1)^{\p{B}+1}(\d_A\chi_B-(-1)^{\p{A}\p{B}}\d_B\chi_A)\,.
\end{equation}
Let us also explicitly write down $d\sigma=0$ for a $2$-form $\sigma$:
\begin{equation}
 \d_A\sigma_{BC}(-1)^{|A||C|+|B|}+\text{cycle}(A,B,C)=0\,.
\end{equation}

%

The contraction of a vector field and a differential form is itself the following vector field on $\Pi T\manM$:
\begin{equation}
\label{inner}
 i_V=V^A\dl{(dz^A)}
\end{equation}
In particular, the components can be expressed as follows
\begin{equation}
  \alpha_{A_1\ldots A_p}(z)= i_{A_1}\ldots i_{A_p}\alpha(z,dz)\,, \quad i_A:=i_{{\frac{\d}{\d z^A}}}\,.
\end{equation}

It is useful to define the Lie derivative as acting on forms through the Cartan formula
\begin{equation}
\label{superlie}
\Lie_V=\commut{i_V}{\dm}=i_V \dm +(-1)^{\p{V}}\dm i_V\,,
\end{equation}
where $\commut{\cdot}{\cdot}$ denotes the graded commutator. $\Lie_V$ is a vector field on $\Pi T\manM$.
Note also the following relations $\commut{\Lie_V}{\dm}=0$. In particular, for functions one gets $\Lie_Vf=V^A \d_Af$.
The action on differential $p$-forms is easily found using the Leibnitz rule and the action on basic differentials:
\begin{equation}
(-1)^{\p{V}}\Lie_V \dm z^A=\dm i_V \dm z^A=\dm (Vz^A)=\dm V^A=dz^B\d_B V^A\,.
\end{equation}
For instance, for $1$-forms one gets
\begin{equation}
\label{liechi}
(-1)^{\p{V}}\Lie_V \chi = dz^A\, \big(\d_A V^B \chi_B + (-)^{|A||V|}\, V^B \d_B \chi_A\big)\;.
\end{equation}

\section*{Acknowledgments}
\label{sec:acknowledgements}

We are indebted to I.~Batalin for the collaboration at the early stage of this project. Useful discussions
with  G.~Barnich, S.~Lyakhovich, R.~Metsaev, M.~Vasiliev and A.~Verbovetsky are gratefully acknowledged.
The work of K.A. was supported by RFBR grant 11-01-00830.  The work of M.G. was supported by RFBR grant 13-01-00386.

\small
\addtolength{\baselineskip}{-3pt}
\addtolength{\parskip}{-3pt}


\providecommand{\href}[2]{#2}\begingroup\raggedright\endgroup

\end{document}